\documentclass[preprint,preprintnumbers,amsmath,amssymb]{revtex4-1}
\usepackage{amsfonts}    
\usepackage{amssymb}
\usepackage{amsmath}
\usepackage{latexsym}
\usepackage{eepic}
\usepackage{graphicx}
\usepackage[usenames,dvipsnames]{color}
\usepackage{color}

\usepackage[active]{srcltx}
\usepackage{epstopdf}

\newcommand{\al}{\alpha}
\newcommand{\pa}{\partial}

\newcommand{\si}{\sigma}
\newcommand{\la}{\lambda}
\newcommand{\ta}{\tau}

\newcommand{\Om}{\Omega}
\newcommand{\om}{\omega}
\newcommand{\de}{\delta}

\newcommand{\De}{\Delta}

\newcommand{\tha}{\theta}

\newcommand{\rar}{\rightarrow}
\newcommand{\lrar}{\leftrightarrow}

\newcommand{\non}{\nonumber}

\begin{document}

\title{Three-body problem in $d$-dimensional space: ground state, (quasi)-exact-solvability.}

\author{Alexander V Turbiner\\[8pt]
Instituto de Ciencias Nucleares, UNAM, M\'exico DF 04510, Mexico\\[8pt]
turbiner@nucleares.unam.mx\\[8pt]
     Willard Miller, Jr.\\[8pt]
School of Mathematics, University of Minnesota, \\
Minneapolis, Minnesota, U.S.A.\\[8pt]
miller@ima.umn.edu\\
[10pt]
and \\[10pt]
M.A.~Escobar-Ruiz,\\[8pt]
School of Mathematics, University of Minnesota, \\
Minneapolis, Minnesota, U.S.A.\\[8pt]
and\\
Centre de Recherches Math\'ematiques, Universit\'e de Montreal, \\
C.P. 6128, succ. Centre-Ville, Montr\'eal, QC H3C 3J7, Canada\\[8pt]
mauricio.escobar@nucleares.unam.mx}

\begin{abstract}
As a straightforward generalization and extension of our previous paper, {\it J. Phys. \bf A50} (2017) 215201 \cite{TMA:2016} we study
aspects of the quantum and classical  dynamics of a $3$-body system with equal masses, each body with $d$ degrees of freedom, with
interaction depending only on mutual (relative) distances. The study is restricted to solutions in the space of relative motion which
are functions of mutual (relative) distances only. It is shown that the ground state (and some other states) in the quantum case and the
planar trajectories (which are in the interaction plane) in the classical case are of this type.
The quantum (and classical) Hamiltonian for which the states are defined by this type eigenfunctions 
is derived. It corresponds to a
three-dimensional quantum particle moving in a curved space with special $d$-dimension-independent metric in a certain $d$-dependent
singular potential, while at $d=1$ it {\it elegantly} degenerates to a two-dimensional particle moving in flat space.
It admits a description in terms of pure geometrical characteristics of the interaction triangle which is defined by the three relative distances.
The kinetic energy of the system is $d$-independent, it has a hidden $sl(4,R)$ Lie (Poisson) algebra structure, alternatively,
the hidden algebra $h^{(3)}$ typical for the $H_3$ Calogero model as in the $d=3$ case. We find an exactly-solvable three-body $S^3$-permutationally
invariant, generalized harmonic oscillator-type potential as well as a quasi-exactly-solvable three-body sextic polynomial type potential
with singular terms. For both models an extra first order integral exists. For $d=1$ the whole family of 3-body (two-dimensional)
Calogero-Moser-Sutherland systems as well as the TTW model are reproduced. It is shown that a straightforward generalization of the 3-body
(rational) Calogero model to $d>1$ leads to two primitive quasi-exactly-solvable problems.
The extension to the case of non-equal masses is straightforward and is briefly discussed.

\end{abstract}

\maketitle

\newpage

\section*{Introduction}

Let us take three classical particles in $d$-dimensional space with potential depending on mutual distances alone. Assume that all the initial
particle positions lie on the same plane. If initial velocities are chosen parallel to this plane then the motion will be planar: all trajectories
are in the plane. Thus, after separation of the center-of-mass motion the trajectories are defined by evolution of the relative (mutual) distances.
The old question is to find equations for trajectories which depend on relative distances only. The aim of the present paper is to construct
the Hamiltonian which depends on relative distances and also  describes such a planar dynamics for the three particle case.
Our strategy is to study the quantum problem and then take the classical limit, making {\it de-quantization}.

The quantum Hamiltonian for three $d$-dimensional particles with translation-invariant potential,
which depends on relative (mutual) distances between particles only, is of the form,
\begin{equation}
\label{Hgen}
   {\cal H}\ =\ -\sum_{i=1}^3 \frac{1}{2 m_i} \De_i^{(d)}\ +\  V(r_{12},\,r_{13},\,r_{23})\ ,\
\end{equation}
with coordinate vector of $i$th particle ${\bf r}_i \equiv {\bf r}^{(d)}_i=(x_{i,1}\,, x_{i,2}\,,x_{i,3}\ldots \,,x_{i,d})$\,, where
\begin{equation}
\label{rel-coord}
r_{ij}=|{\bf r}_i - {\bf r}_j|\ ,\quad i,j=1,2,3\ ,
\end{equation}
is the (relative) distance between particles $i$ and $j$, $r_{ij}=r_{ji}$,
sometimes called the Jacobi distances.

\begin{figure}[h]
  \centering
  \includegraphics[width=13.0cm]{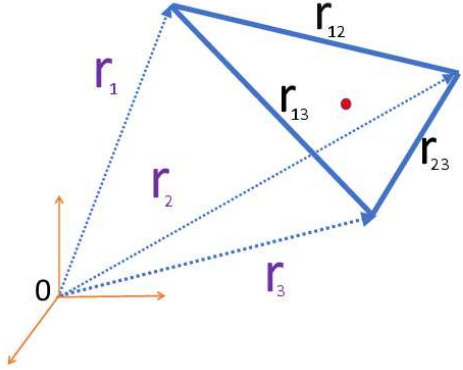}
  \caption{Three-particle system: the coordinate vectors ${\bf r}_i$ mark positions of vertices of the triangle of interaction with sides $r_{ij}$. The center-of-mass (the barycenter of the triangle) is marked by a (red) bubble.}
\label{Fig}
\end{figure}

The number of relative distances is equal to the number of edges of the triangle which is
formed by taking the particle positions as vertices. We call this triangle the {\it triangle of interaction},
see for illustration Fig.\ref{Fig}. Here, $\De_i^{(d)}$ is the $d$-dimensional Laplacian,
\[
     \De_i^{(d)}\ =\ \frac{\pa^2}{\pa{{\bf r}_i} \pa{{\bf r}_i}}\ ,
\]
associated with the $i$th body.
For simplicity all masses in (\ref{Hgen}) are assumed to be equal: $m_i=m=1$. The configuration space
for ${\cal H}$ is ${\bf R}^{3d}$.
The center-of-mass motion described by vectorial coordinate
\[
    {\bf R}_{_0} \ =\ \frac{1}{{\sqrt 3}}\,\sum_{k=1}^{3} {\bf r}_{_k}\ ,
\]
can be separated out; this motion is described by a $d$-dimensional plane wave, $\sim e^{i\, {\bf k} \cdot {\bf R}_{_0}}$.

The spectral problem is formulated in the space of relative motion ${\bf R}_r \equiv {\bf R}^{2d}$; it is of the form,
\begin{equation}
\label{Hrel}
   {\cal H}_r\,\Psi(x)\ \equiv \ \bigg(- \frac{1}{2}\,\De_r^{(2d)} + V(r_{12},\,r_{13},\,r_{23})\bigg)\, \Psi(x)\ =\ E \,\Psi(x)\ ,\
   \Psi \in L_2 ({\bf R}_r)\ ,
\end{equation}
where $\De_r^{(2d)}$ is the flat-space Laplacian in the space of relative motion.
If the space of relative motion ${\bf R}_r$ is parameterized by two, $d$-dimensional vectorial Jacobi coordinates
\begin{equation}
\label{rj}
     {\bf q}_{j} \ = \ \frac{1}{\sqrt{j(j+1)}}\sum_{k=1}^j k\,({\bf r}_{k+1} - {\bf r}_{{k}})\ ,
        \qquad\qquad j=1,2\ ,
\end{equation}
the flat-space $2d$-dimensional Laplacian in the space of relative motion becomes diagonal
\begin{equation}
\label{Dflat}
       \De_r^{(2d)}\ =\ \sum_{j=1,2} \frac{\pa^2}{\pa{{\bf q}_j} \pa{{\bf q}_j}}\ .
\end{equation}
Thus, ${\bf q}_{j}$ plays a role of the Cartesian coordinate vector in the space of relative motion.

The case $d=1$ (three bodies on a line) is special. The triangle of interaction degenerates into an interval with the marked point inside -
the vertices of the triangle correspond to the endpoints and the marked point - its area is equal to zero. Moreover, for $d=1$ the relative distances obey the constraint (hyperplane condition),
\begin{equation}
\label{rel3}
   r_{12} + r_{23} +  r_{13}\ =\ 0\ ,
\end{equation}
where it is assumed that $r_{13}=-r_{31}$.
Hence, the three relative distances are related and only two relative distances can serve as independent variables. Therefore, the Laplacian in the space of relative variables becomes
\begin{equation}
\label{Drel3-1}
   \De_r^{(2)}\ =\ 2\,\bigg(\frac{\pa^{2}}{\pa r_{12}^{2}}\ +\
   \frac{\pa^{2}}{\pa r_{23}^{2}} - \frac{\pa^{2}}{\pa r_{12} \,\pa r_{23}}\bigg) \ ,
\end{equation}
cf. (\ref{Dflat}). The configuration space ${\bf R}_r$ is the quadrant, $r_{12}, r_{23} \geq 0$.

\bigskip

{\large \it Observation} \cite{Ter}\,:
\begin{quote}
  There exists a family of  eigenstates of the Hamiltonian (\ref{Hgen}), including the ground state,
  which depend on the three relative distances $\{r_{ij}\}$ alone\,. The same is correct for the $n$ body problem:
  there exists a family of eigenstates, including the ground state, which depend on  relative distances alone.
\end{quote}


This observation is presented for the case of scalar particles, bosons. It can be generalized to the case of fermions,
\bigskip

{\large \it Conjecture}\,:
\begin{quote}
   In the case of three fermions there exists a family of the eigenstates of the Hamiltonian (\ref{Hgen}), including the ground state, in which the coordinate functions depend on three relative distances $\{r_{ij}\}$ only\,. The same is correct for the $n$ body problem:
   there exists a family of the eigenstates, including the ground state, in which the coordinate functions depend on relative distances only.
\end{quote}

\bigskip

\noindent
Our primary goal is to find the differential operator in the space of relative distances $\{r_{ij}\}$ for which these states are eigenstates. In other words, to find a differential equation depending only on $\{r_{ij}\}$ for which these states are solutions. This  implies a study of the evolution of the triangle of interaction with fixed barycenter (center-of-mass). We consider the case of three spinless particles.

In our previous paper \cite{TMA:2016} the physically important case $d=3$ was studied in detail. In this paper it will be shown that the generalization to arbitrary $d$ is straightforward. Except for $d=1$ almost all formulas remain conceptually unchanged, the presentation (and even wording) remains almost the same and most conclusions are unaltered.


\section{Generalities}

As a first step let us change variables in the space of relative motion ${\bf R}_r$:
$$({\bf q}_{j}) \lrar (r_{ij}, \Om)\ ,$$
where for $d>1$ the number of (independent) relative distances $r_{ij}$ is equal to 3 and $\Om$ is a collection of $(2d-3)$ angular variables. Thus, we split ${\bf R}_r$ into a combination of the space of relative distances ${\bf \tilde R}$ and a space parameterized by angular variables, essentially those on the  sphere $S^{2d-3}$. There are known several ways to
introduce variables in ${\bf R}_r$: the perimetric  coordinates by Hylleraas \cite{Hylleraas}, the scalar products of vectorial Jacobi
coordinates ${\bf r}_{ij}$ \cite{Gu} and the three relative (mutual) distances $r_{ij}$ (see e.g. \cite{Loos}). We follow the last one.
In turn, the angular variables are introduced as the Euler angles on the $S^{2d-4}$ sphere defining the normal to the interaction plane (triangle) and the azimuthal angle of rotation of the interaction triangle around its barycenter, see e.g. \cite{Gu}.

A key observation is that in new coordinates $(r_{ij}, \Om)$ the flat-space Laplace operator (the kinetic energy operator) in the space of
relative motion ${\bf R}_r$ takes the form of the sum of two second-order differential operators
\begin{equation}
\label{addition}
    \frac{1}{2}\De_r^{(2d)}\ =\ {\De_R}(r_{ij}, \pa_{ij}) + {\tilde \De} (r_{ij}, \Om, \pa_{ij}, \pa_{\Om})
    \ ,\quad \pa_{ij} \equiv \frac{\pa}{\pa r_{ij}}\ ,
\end{equation}
where the first operator depends on relative distances {\it only}, (hence, the coefficient functions do not depend on angles),
while the second operator depends on angular derivatives in such a way that it annihilates any angle-independent function,
\[
  {\tilde \De} (r_{ij}, \Om, \pa_{ij}, \pa_{\Om})\, \Psi(r_{ij})\ =\ 0\ .
\]
This observation holds for the $n$-body case as well \cite{MTA:2017}.

For $d=1$ the operator ${\tilde \De}$ is absent (no angular variables occur) and we have
\[
  {\De_R}(r_{ij}, \pa_{ij})\ =\ \De_r^{(2)}\ ,
\]
see (\ref{Drel3-1}).
In general, for $d>1$ the commutator $$[{\De_R}(r_{ij})\ ,\ {\tilde \De} (r_{ij}, \Om, \pa_{\Om})] \neq 0\ .$$

If we look for angle-independent solutions of (\ref{Hrel}), due to the decomposition (\ref{addition}) the general spectral problem (\ref{Hrel}) reduces to a particular 3-body spectral problem
\begin{equation}
\label{Hrel-Mod}
   {\tilde {\cal H}}_R\,\Psi(r_{ij})\ \equiv \ \bigg(- {\De_R}(r_{ij}, \pa_{ij}) + V(r_{ij})\bigg)\, \Psi(r_{ij})\ =\ E\,\Psi(r_{ij})\ ,\quad
   \Psi \in L_2 ({\bf \tilde R})\ ,
\end{equation}
where ${\bf \tilde R}$ is the space of relative distances. Clearly, we can write
\begin{equation}
\label{gmunu}
     {\De_R}(r_{ij}, \pa_{ij})\ =\ g^{\mu \nu}(r) \pa_{\mu} \pa_{\nu}\ +\ b^{\mu} \pa_{\mu}\ ,
\end{equation}
where $g^{\mu \nu}(r)$ is the matrix made out of coefficients in front of the second derivatives and $b^{\mu}$ is a column vector.

Surprisingly, for any $d>1$ one can find the $d$-dependent gauge factor $\Gamma(r_{ij})$ such that the operator ${\De_R}(r_{ij}, \pa_{ij})$ takes the form of the Schr\"odinger operator,
\begin{equation}
\label{DLB}
     \Gamma^{-1}\,{\De_R}\,(r_{ij}, \pa_{ij})\, \Gamma\ =\ {\De_{LB}}(r) - {\tilde V}(r)\ \equiv
      -{\tilde \De}_R \ ,
\end{equation}
where $\De_{LB}(r)$ is the Laplace-Beltrami operator with contravariant, $d$-independent metric
$g^{\mu \nu}(r)$, in general, on some non-flat, (non-constant curvature) manifold. It makes sense of the kinetic energy. Here ${\tilde V}(r)$ is the effective potential. The potential ${\tilde V}$ becomes singular at the boundary of the configuration space, where the determinant $D(r)=\det g^{\mu \nu}(r)$ vanishes. The operator ${\tilde \De}_R$ is Hermitian with measure $D(r)^{-\frac{1}{2}}$. Eventually, we arrive at the spectral problem for the Hamiltonian
\begin{equation}
\label{Hrel-final}
   {H}_{LB}(r)\ \equiv\ -{\De_{LB}}(r) + V(r) + {\tilde V}(r)\ ,
\end{equation}
with $d$-independent kinetic energy ${\De_{LB}}(r)$. Again the case $d=1$ is special, the gauge factor is trivial, $\Gamma=1$,
and
\[
     {\De_{LB}}(r) \ =\ {\De_R}(r)\ =\ \De_r^{(2)}\ .
\]

Following the {\it de-quantization} procedure of replacement of the quantum momentum (derivative) by the classical momentum
\[
      -i\,\pa\ \rar\ P\ ,
\]
one can get a classical analogue of (\ref{Hrel-final}),
\begin{equation}
\label{Hrel-Cl-final}
   {H}^{(c)}_{LB}\ =\ g^{\mu \nu}(r) P_{\mu} P_{\nu} + V(r) + {\tilde V}(r)\ .
\end{equation}
It describes the internal motion of a 3-dimensional body with tensor of inertia $(g^{\mu \nu})^{-1}$\, with center of mass  fixed.

The Hamiltonians (\ref{Hrel-final}), (\ref{Hrel-Cl-final}) are the main objects of study of this paper.

\section{Three-body case: $d=1$, concrete results }


In the one-dimensional case $d=1$ the Laplace-Beltrami operator in (\ref{Hrel-final}) becomes
\[
  {\De_{LB}}(r)\ =\ 2\bigg(\frac{\pa^{2}}{\pa r_{12}^{2}}\ +\
   \frac{\pa^{2}}{\pa r_{23}^{2}} - \frac{\pa^{2}}{\pa r_{12} \,\pa r_{23}}\bigg) \ ,
\]
see (\ref{Drel3-1}). It  corresponds to the two-dimensional flat space Laplacian and is evidently an algebraic operator.
Formally, it is not $S_3$ invariant unlike the original $3d$-Laplacian in (\ref{Hgen}),
the kinetic energy, although it remains $S_2$ invariant. Note that in variables,
\[
r_{12}^2\ =\ \rho_{12}\ ,\ r_{13}^2\ =\ \rho_{13}\ ,\ r_{23}^2\ =\ \rho_{23}\ ,
\]
see below, the emerging flat-space Laplacian ${\De_{LB}}(\rho)$ is not anymore algebraic.
As a realization of $S_2$ invariance of (\ref{Drel3-1}) let us introduce $S_2$ invariants
\begin{equation}
\label{d1-xi}
 \xi_1 = r_{12} + r_{23}\quad ,\quad \xi_2 = r_{12} \, r_{23}\ ,
\end{equation}
as new variables, which is a polynomial change of variables, then
\begin{equation}
\label{Drel3-1-xi}
  {\De_{LB}}(\xi)\ =\ 2\,\bigg(\frac{\pa^{2}}{\pa \xi_{1}^{2}}\ +\ (\xi_{1}^2 - 3 \xi_{2})
   \frac{\pa^{2}}{\pa \xi_{2}^{2}} + \xi_{1} \frac{\pa^{2}}{\pa \xi_{1} \,\pa \xi_{2}}  - \frac{\pa}{\pa \xi_{2}} \bigg) \ .
\end{equation}
This  is an algebraic operator which can be rewritten in terms of the
generators of the maximal affine subalgebra $b_2$ of the algebra $sl(3,{\bf R})$ in $\xi$-variables, c.f. below (\ref{sl4R}), see  \cite{RT:1995,Turbiner:1998}.

There exists another polynomial $S_2$-symmetric change of variables \cite{Turbiner:1998}
\begin{equation}
\label{d1-si}
 \si_2 = - r_{12} r_{23} - r_{12}^2 - r_{23}^2\ =\ \xi_2 - \xi_1^2\ ,\
 \si_3 = - r_{12} \, r_{23}(r_{12} + r_{23}) = - \xi_1 \xi_2\ ,
\end{equation}
which leaves the operator ${\De_{LB}}$ algebraic,
\begin{equation}
\label{Drel3-1-si}
  {\De_{LB}}(\si)\ =\ -2\,\bigg(3 \si_2\,\frac{\pa^{2}}{\pa \si_{2}^{2}}\ -\
      \si_{2}^2 \frac{\pa^{2}}{\pa \si_{3}^{2}} + 9\, \si_{3} \frac{\pa^{2}}{\pa \si_{2} \,\pa \si_{3}}
      + 3 \frac{\pa}{\pa \si_{2}} \bigg) \ .
\end{equation}
In fact, the variables (\ref{d1-si})
\begin{equation}
\label{d-si}
   \si_1 = r_{12} + r_{23} + r_{31}\ ,\
   \si_2 = r_{12} r_{23} + r_{12} r_{13} + r_{23} r_{13}\quad ,\quad \si_3 = r_{12} \, r_{23} \, r_{13}\ ,
\end{equation}
are $S_3$ invariants, subject to the condition that the 1st invariant $\si_1$ vanishes,
\[
    \si_1\ =\ 0\ .
\]
Hence, the operator (\ref{Drel3-1-si}) is, in fact, $S_3$ permutationally invariant. It can be immediately seen that
the operator (\ref{Drel3-1-si}) describes the kinetic energy of  relative motion of the 3-body $(A_2)$ rational Calogero model
\cite{Calogero:1969} with potential
\begin{equation}
\label{Vcalogero}
    V_{A_2}\ =\ g \bigg(\frac{1}{r_{12}^2} + \frac{1}{r_{23}^2} + \frac{1}{r_{13}^2}\bigg)\ =\
    g \bigg(\frac{1}{\rho_{12}} + \frac{1}{\rho_{23}} + \frac{1}{\rho_{13}}\bigg)\ ,
\end{equation}
in  algebraic form, see \cite{Turbiner:1998}. It is easy to check that the potential $V_{A_2}$ is a rational function
in $\si_{2,3}$ (\ref{d1-si}). It was shown in \cite{RT:1995,Turbiner:1998} that the Hamiltonian of  relative motion of the 3-body $(A_2)$ rational
Calogero model (even with potential modified by adding the harmonic oscillator potential), gauge-rotated
with its ground state function and written in variables $\si_{2,3}$ is an algebraic operator as well. This operator
can be rewritten in terms of the generators of the maximal affine subalgebra $b_2$ of the algebra
$sl(3,{\bf R})$
\begin{eqnarray}
\label{sl3R}
 {\cal J}_i^- &=& \frac{\pa}{\pa u_i}\ ,\qquad \quad i=1,2\ , \non  \\
 {{\cal J}_{ij}}^0 &=&
               u_i \frac{\pa}{\pa u_j}\ , \qquad i,j=1,2 \ , \\
 {\cal J}^0(N) &=& \sum_{i=1}^{2} u_i \frac{\pa}{\pa u_i}-N\, , \non \\
 {\cal J}_i^+(N) &=& u_i {\cal J}^0(N)\ =\
    u_i\, \left( \sum_{j=1}^{2} u_j\frac{\pa}{\pa u_j}-N \right)\ ,
       \quad i=1,2\ ,
\end{eqnarray}
where $N$ is a parameter and
\[
    u_1\ =\ \si_2\ ,\ u_2\ =\ \si_3\ .
\]

Another polynomial change of variables
\begin{equation}
\label{d1-la}
 \la_1 = \si_2 = - r_{12} r_{23} - r_{12}^2 - r_{23}^2\ =\ \xi_2 - \xi_1^2\ ,\
 \la_2 = \si_3^2 = r_{12}^2 \, r_{23}^2(r_{12} + r_{23})^2 = \xi_1^2 \xi_2^2\ ,
\end{equation}
leaves the operator ${\De_{LB}}$ algebraic,
\begin{equation}
\label{Drel3-1-la}
  {\De_{LB}}(\la)\ =\ -2\,\bigg(3 \la_1\,\frac{\pa^{2}}{\pa \la_{1}^{2}}\ -\
      4\la_{1}^2 \la_2 \frac{\pa^{2}}{\pa \la_{2}^{2}} + 18\, \la_{2} \frac{\pa^{2}}{\pa \la_{1} \,\pa \la_{2}}
       + 3 \frac{\pa}{\pa \la_{1}} - 2 \la_{1}^2 \frac{\pa}{\pa \la_{2}}  \bigg) \ .
\end{equation}
It can be immediately seen that the operator (\ref{Drel3-1-la}) describes the kinetic energy of relative motion of the 3-body $(G_2)$ rational Wolfes model \cite{Wolfes:1974}
with potential
\[
    V_{G_2}\ =\ g \bigg(\frac{1}{r_{12}^2} + \frac{1}{r_{23}^2} + \frac{1}{r_{13}^2}\bigg)\ +\
    g_1 \bigg(\frac{1}{(r_{12}-r_{23})^2} + \frac{1}{(r_{23}-r_{31})^2} + \frac{1}{(r_{12}-r_{31})^2}\bigg)\ ,
\]
in  algebraic form, see \cite{Turbiner:1998}. It can be rewritten in terms of the
generators of the algebra $g^{(2)}$, see \cite{Turbiner:1998}. It is easy to check that the potential $V_{G_2}$ is a rational
function in $\la_{1,2}$ (\ref{d1-la}). It was shown in
\cite{RT:1995,Turbiner:1998} that the Hamiltonian of  relative motion of the 3-body $(G_2)$
rational Wolfes model (even with potential modified by adding the harmonic oscillator potential), gauge-rotated with its ground state
function and written in variables $\la_{1,2}$ is an algebraic operator.
This operator can also be rewritten in terms of the generators of the algebra $g^{(2)}$.

The most general polynomial change of variables known so far, which leaves the operator ${\De_{LB}}$ algebraic, has led to the discovery of
the so-called TTW model \cite{TTW:2009} - the most general superintegrable and exactly-solvable model on the plane.
Following Calogero \cite{Calogero:1969} let us introduce polar coordinates in the space of relative distances,
\[
 q_1\,=\,\frac{1}{\sqrt{2}}\,r_{12} = r \cos \varphi\ ,\  q_2\,=\,\sqrt{\frac{2}{3}}\,(r_{13}+r_{23}) = r \sin \varphi\ ,
\]
see (\ref{rj}). Now we define new variables
\[
t\ =\ r^2\ ,\ u\ =\ r^{2k} \sin^2 {k\varphi}\ ,
\]
which are the invariants of the dihedral group $I_2(k)$ for integer $k$\, (and even rational $k$). The operator (\ref{Drel3-1}) takes an amazingly simple algebraic form,

\[
  {\De_{LB}}^{(k)} \ =\  -4 t \pa^2_t - 8k u \pa^2_{tu} - 4k^2 t^{k-1} u \pa^2_u
    - 4\pa_t  - 2 k^2 t^{k-1}\pa_u \ ,
\]
for $k=1,2,\ldots$\,. This operator can be rewritten in terms of the generators of the algebra $g^{(k)}$,
see \cite{TTW:2009}\,. Forming the Schr\"odinger operator
\begin{equation}
\label{TTW}
  {\cal H}_{TTW}^{(k)}\ =\ -{\De_{LB}}^{(k)} + \om^2 t + \frac{k^2\al\, t^{k-1}}{t^k-u} + \frac{k^2 \beta\, t^{k-1}}{u}\ ,
\end{equation}
we arrive at the Tremblay-Turbiner-Winternitz (TTW) model; if $k=3$ at $\al=0$ the 3-body $(A_2)$ rational Calogero model occurs, otherwise for $\al \neq 0$ the 3-body $(G_2)$ rational Wolfes model occurs, both after separation out center-of-mass motion. Interestingly, for $k=1$ we get the Smorodinsky-Winternitz model, while for $k=2$ it will be the $BC_2$ rational model. Both models describe the relative motion of 3-body problem.

The TTW model is exactly-solvable and integrable for any real $k$ - there exists a 2nd order integral (the 1st order symmetry operator squared),
while for rational $k=p/q$ the model is superintegrable - there exists an integral of  order $2(p+q)-1$ \cite{KKM:2010}.
For integer $k$ by gauge-rotation of the Hamiltonian (\ref{TTW}) with its ground state function one can transform it to an algebraic operator.
The same is true for both integrals: gauge rotation with ground state eigenfunction leads them to a form of algebraic operators in variables
$t,u$\,. All these algebraic operators can be rewritten in terms of the generators of the algebra $g^{(k)}$.

It must be emphasized that there are  non-polynomial changes of variables $(r_{12}\,,\, r_{23})$ (trigonometric and elliptic) in (
\ref{Drel3-1}) which can still lead to algebraic operators. In the case of trigonometric change of variables there occurs the kinetic energy
of the relative motion of 3-body, $(A_2)$ trigonometric Sutherland model \cite{RT:1995}, or $(G_2)$ trigonometric Sutherland model
\cite{Turbiner:1998}, or the kinetic energy of $BC_2$ trigonometric model \cite{Brink:1997}. For discussion see \cite{Turbiner:2013}.

In the case of elliptic change of variables there occurs the kinetic energy of  relative motion of the 3-body, $(A_2)$ elliptic
Calogero model, or $(G_2)$ elliptic model \cite{ST:2015}, or the kinetic energy of the  $BC_2$ elliptic model \cite{DGU:2001}.

\section{Three-body case: $d > 1$, concrete results }

\subsection{$r$-representation}

Assuming $d > 1$, after straightforward calculations the operator ${\De_R}(r_{ij}, \pa_{ij})$ (in decomposition (\ref{addition})) can be found to be
\[
   2{\De_R}(r_{ij}, \pa_{ij})\ =\ \bigg[\ 2\,(\pa^{2}_{r_{12}} +\pa^{2}_{r_{23}}+\pa^{2}_{r_{13}})
+ \frac{2(d-1)}{r_{12}}\,\pa_{r_{12}}  +  \frac{2(d-1)}{r_{23}}\,\pa_{r_{23}} + \frac{2(d-1)}{r_{13}}\,\pa_{r_{13}}
\]
\begin{equation}
\label{addition3-3r}
  + \frac{r_{12}^2-r_{13}^2+r_{23}^2}{r_{12} r_{23}}\,\pa_{r_{12}}\pa_{r_{23}}  + \frac{r_{12}^2+r_{13}^2-r_{23}^2}{r_{12} r_{13}}\,\pa_{r_{12}}\pa_{r_{13}} +  \frac{r_{13}^2+r_{23}^2-r_{12}^2}{r_{13} r_{23}}\,\pa_{r_{23}}\pa_{r_{13}} \ \bigg]\ ,
\end{equation}
cf. e.g. \cite{Loos}. Note that at $d=1$ the operator (\ref{addition3-3r}) becomes degenerate. This can be seen by calculating the determinant of
the metric $g^{\mu \nu}$, see (\ref{gmunu}) for definition, of the operator (\ref{addition3-3r}),
\[
D(r)=\det g^{\mu \nu}(r)\ =\  \frac{12}{r_{12}^2\,r_{23}^2\,r_{13}^2}(r_{12}^2+r_{23}^2+r_{13}^2)\,S^2_{\triangle} \ ,
\]
where $S_{\triangle}$ is the area of the interaction triangle, see below. For $d=1$ the interaction triangle shrinks to the interval with marked point,
its area vanishes, $S_{\triangle}=0$, and the determinant is identically zero, $D(r)=0$. This implies that the original three-dimensional
configuration space at $d>1$ given by $S_{\triangle}\geq 0$, shrinks to the boundary, $S_{\triangle}=0$ and effectively becomes two-dimensional.

In general, the operator (\ref{addition3-3r}) does not depend on the choice of the angular variables $\Om$. While the operator
${\tilde \De} (r_{ij}, \pa_{ij}, \Om, \pa_{\Om})$ in (\ref{addition}), for example at $d=2$, where there is a single angular variable,
$\Om=\tha$, is equal to
\[
{\tilde \De} \ =\
 -\ \frac{12\,S_{\triangle}}{r_{12}^2\,r_{23}^2\,r_{13}^2}\,\bigg({r_{12}^3}\,\pa_{r_{12}}\, +\, {r_{23}^3}\,\pa_{r_{23}}\, +\,{r_{13}^3}\,\pa_{r_{13}}\,
  -\,\frac{r_{12}^4+r_{23}^4+r_{13}^4}{S_{\triangle}}\,\pa_{\tha} \bigg)\pa_{\tha}\ ,
\]
where $\tha$ is the azimuthal angle of rotation around barycenter of triangle of interaction, see Fig.~1 and $S_{\triangle}$ is the area of
the interaction triangle, see below Eq.(\ref{CFrho}). It is evident that ${\tilde \De}$ annihilates any angle-independent function.
The variable $\tha$ is not separated in ${\tilde \De}$ and, thus, in $\De_r^{(2d)}$ (\ref{addition}): the eigenfunctions in (\ref{Hrel}) are
{\it not} factorizable to the form $R(r_{ij})\, A (\tha)$.

The configuration space in the space of relative distances is
\begin{equation}
\label{CFr}
 0 < r_{12},r_{13},r_{23} < \infty\,,\quad  r_{23} < \ r_{12} + r_{13}\,,\quad r_{13}< r_{12}+r_{23}\,,\quad r_{12}< r_{13}+r_{23}\ ,
\end{equation}
 equivalent to $S_{\triangle}>0$.
In the space with Cartesian coordinates $(x,y,z)=(r_{12},r_{13},r_{23})$ the configuration space lies in the first octant and is the interior of the inverted  tetrahedral-shaped object with base at infinity, vertex at the origin and edges $(t,t,2t)$, $(t,2t,t)$ and $(2t,t,t)$, $0\leq t<\infty$.

\subsection{$\rho$-representation}

Formally, the operator (\ref{addition3-3r}) is invariant under reflections $Z_2 \oplus Z_2 \oplus Z_2$,
\[
r_{12} \rightarrow -r_{12}  \ ,\qquad  r_{13} \Leftrightarrow -r_{13} \ ,\qquad  r_{23} \Leftrightarrow -r_{23} \ ,
\]
and w.r.t. $S_3$-group action. If we introduce new variables,
\begin{equation}
\label{rho}
r_{12}^2\ =\ \rho_{12}\ ,\ r_{13}^2\ =\ \rho_{13}\ ,\ r_{23}^2\ =\ \rho_{23}\ ,
\end{equation}
the operator (\ref{addition3-3r}) becomes algebraic,
\[
  {\De_R}(\rho)\ =\ 4(\rho_{12} \pa^2_{\rho_{12}} + \rho_{13} \pa^2_{\rho_{13}} +\rho_{23} \pa^2_{\rho_{23}})\ +\
\]
\[
  2 \bigg((\rho_{12} + \rho_{13} - \rho_{23})\pa_{\rho_{12}}\pa_{\rho_{13}}\ +
          (\rho_{12} + \rho_{23} - \rho_{13})\pa_{\rho_{12}}\pa_{\rho_{23}}\ +
          (\rho_{13} + \rho_{23} - \rho_{12})\pa_{\rho_{13}}\pa_{\rho_{23}}
    \bigg)\  +
\]
\begin{equation}
  2d (\pa_{\rho_{12}} + \pa_{\rho_{13}}+ \pa_{\rho_{23}}) \ ,
\label{addition3-3rho}
\end{equation}
c.f. \cite{TMA:2016} at $d=3$.
Note that the operator (\ref{addition3-3rho}) is of Lie-algebraic nature: it can be rewritten in terms of the generators of the algebra
$sl(4,{\bf R})$ realized by the first order differential operators, see below. It acts as a filtration for the flag of finite-dimensional
representation spaces of this algebra
\[
     {\mathcal P}^{(1,2,3)}_{N}\ =\ \langle \rho_{12}^{p_1} \rho_{13}^{p_2} \rho_{23}^{p_3} \vert \
     0 \le p_1 + p_2+ p_3 \le N \rangle\ ,\ N=0,1,2, \ldots\ .
\]

From (\ref{CFr}) and (\ref{rho}) it follows that the corresponding configuration space in $\rho$ variables is given by the conditions
\[
0 < \rho_{12},\rho_{13},\rho_{23} < \infty\ ,\
{\rho}_{23} <  (\sqrt{{\rho}_{12}} + \sqrt{{\rho}_{13}})^2,\ {\rho}_{13} < (\sqrt{{\rho}_{12}} + \sqrt{{\rho}_{23}})^2,\ {\rho}_{12} < \ (\sqrt{{\rho}_{13}} + \sqrt{{\rho}_{23}})^2\ ,
\]
 equivalent to $S_{\triangle}>0$. The boundary of the configuration space is given by $S(\rho)=S^2_{\triangle}=0$.
We remark that
\begin{equation}
\label{CFrho}
\quad
-16\,S  \ \equiv \  \rho _{12}^2+\rho _{13}^2+\rho _{23}^2 -2 (\rho _{12} \rho _{13}+  \rho _{12} \rho _{23}+
        \rho _{13} \rho _{23}) \ \leq \ 0   \ ,
\end{equation}
because the left-hand side (l.h.s.) is equal to
$$-(r_{12}+r_{13}-r_{23})(r_{12}+r_{23}-r_{13})(r_{13}+r_{23}-r_{12})(r_{12}+r_{13}+r_{23})$$
and conditions (\ref{CFr}) should hold. Therefore, following the Heron formula, l.h.s. is proportional to the square of the area of the triangle of interaction $S^2_{\triangle} = S$\,.

The associated contravariant metric for the operator ${\De_R}(\rho)$ defined by coefficients in front of second derivatives is remarkably simple
\begin{equation}
\label{gmn33-rho}
 g^{\mu \nu}(\rho)\ = \left|
 \begin{array}{ccc}
 4\rho_{12} & \ \rho_{12} + \rho_{13} - \rho_{23} & \ \rho_{12} + \rho_{23} - \rho_{13} \\
            &                                   &                                   \\
 \rho_{12} + \rho_{13} - \rho_{23} & \  4\rho_{13} & \ \rho_{13} + \rho_{23} - \rho_{12} \\
            &  \                                  &                                   \\
 \rho_{12} + \rho_{23} - \rho_{13} & \ \rho_{13} + \rho_{23} - \rho_{12} & 4\rho_{23}
 \end{array}
               \right| \ ,
\end{equation}
it is linear in $\rho$-coordinates(!) with factorized determinant
\begin{equation}
\label{gmn33-rho-det}
\det g^{\mu \nu}(\rho)\ =\ - 6\left(\rho _{12}+\rho _{13}+\rho _{23}\right)
                     \left(\rho _{12}^2+\rho _{13}^2+\rho _{23}^2 -2 (\rho _{12}\, \rho _{13}+
                      \rho _{12}\, \rho _{23}+ \rho _{13} \,\rho _{23})\,\right) \equiv D(\rho) \geq 0\ ,
\end{equation}
and is positive definite. It does not depend on dimension $d$, c.f. \cite{TMA:2016}, explicitly. However, at $d=1$ this determinant vanishes
identically(!) (see below).  It is worth noting a remarkable factorization property of the determinant
\[
D(\rho)\ =\ 6\,(r_{12}^2+r_{13}^2+r_{23}^2)\ \times
\]
\[
(r_{12}+r_{13}-r_{23})(r_{12}+r_{23}-r_{13})(r_{13}+r_{23}-r_{12})(r_{12}+r_{13}+r_{23})\ =
\]
\[
   =\ 96\, P \ S^2_{\triangle}\ ,
\]
where $P=r_{12}^2+r_{13}^2+r_{23}^2$ - the sum of squares
of the sides of the interaction triangle - the trace of metric tensor, $\mbox{Tr}\, g^{\mu \nu}(\rho)=P/4\,$.

The operator (\ref{addition3-3rho}) is $S_3$ permutationally-invariant. Hence, it can be rewritten in terms
of elementary symmetric polynomials $\si_{1,2,3}$,
\begin{equation}
\label{taus}
\begin{aligned}
&  \ta_1\ =  \ \si_1(\rho _{12},\,\rho _{13},\,\rho _{23}) \ = \ \rho _{12}+\rho _{13}+\rho _{23}\ ,
\\ & \ta_2 \ =  \ \si_2(\rho _{12},\,\rho _{13},\,\rho _{23}) \ = \  \rho _{12} \,\rho _{13}+
 \rho _{12} \,\rho _{23}+ \rho _{13}\, \rho _{23} \ ,
\\
&  \ta_3\ = \ \si_3(\rho _{12},\,\rho _{13},\,\rho _{23}) \ = \ \rho _{12}\,\rho _{13}\,\rho _{23}\ .
\end{aligned}
\end{equation}
The determinant $D(\rho)$ has the very simple form,
\begin{equation}
\label{gmn33-rho-det-gamma}
      D(\rho)\ =\ 6\, \ta_1\ (4\ta_2-\ta_1^2)\ ,
\end{equation}
with $$16\, S^2_{\triangle} \ = \ (4\,\ta_2-\ta_1^2)\ ,$$ where only the elementary symmetric polynomials $\ta_{1,2}$ are involved.
When $\det g^{\mu \nu}(\rho)=0$, hence, either $\ta_1=0$, or $\ta_1^2 = 4 \ta_2$, it defines the boundary of the configuration space, see (\ref{CFrho}).

\subsection{$\tau$-representation}

The operator (\ref{addition3-3rho}) being rewritten in terms of elementary symmetric polynomials $\si_{1,2,3}$ in $\rho$-variables (\ref{taus}), remains algebraic

\[
    {\De_R}(\tau) \ = \ 6\,\ta_1\pa_{\ta_1}^2\ +\ 2\ta_1(7\ta_2-\ta_1^2)\pa_{\ta_2}^2\ +\ 2\ta_3(6\ta_2-\ta_1^2)\pa_{\ta_3}^2
   \ +\ 24\,\ta_2\pa_{1,2}^2\ +\ 36\ta_3\pa_{{\ta_1},{\ta_3}}^2
\]
\begin{equation}
\label{addition3-3tau} \ +\ 2\,[9\ta_3\ta_1 +  \ta_2 (4\,\ta_2 - \ta_1^2)]\pa_{{\ta_2},{\ta_3}}^2\ +\
 6\,d\,\pa_{\ta_1}\ +\ 2\,(2d+1)\ta_1\,\pa_{\ta_2}\ +\ 2\,[(d+4)\ta_2-\ta_1^2]\,\pa_{\ta_3}\ ,
\end{equation}
with metric
\begin{equation}
\label{gmn33-tau}
 g^{\mu \nu}(\tau)\ = \left|
 \begin{array}{ccc}
 6\,\ta_1 & 12\,\ta_2 & 18\,\ta_3 \\
            &                                   &                         \\
 12\,\ta_2 & 2\,\ta_1\,(7\,\ta_2-\ta_1^2) &\ 9\,\ta_3\,\ta_1 + 4\,\ta_2^2 - \ta_2 \,\ta_1^2 \\
            &                                   &                         \\
 18\,\ta_3 &\ 9\,\ta_3\,\ta_1 + 4\,\ta_2^2 - \ta_2 \ta_1^2\ & 2\,\ta_3\,(6\,\ta_2-\ta_1^2)
 \end{array}
               \right| \ ,
\end{equation}
see below (\ref{hQES-N-tau}), (\ref{hES-N-tau}). Its determinant,
\[
\det g^{\mu \nu}(\tau)\ =\ 6\, \ta_1\, \left(4\ta_2-\ta_1^2 \right)
  [   2\, \ta_1\, (9\,\ta_2 - 2 \,\ta_1^2)\,\ta_3 -(4\,\ta_2-\ta_1^2) - 27 \,\ta_3^3]
\ ,
\]
c.f. (\ref{gmn33-rho-det-gamma}). Again, the determinant vanishes, if $(16\,S^2_{\triangle})=({4\ta_2-\ta_1^2})=0$.

\subsection{Geometrical variables representation}

It is important to point out  that the operator ${\De_R}(\tau)$ can be rewritten in geometrical terms using them as {\it geometrical} variables
\begin{equation}
\label{G}
\begin{aligned}
&  P \ = \ \ta_1 \  = \  \rho _{12} \ + \ \rho _{13}\ + \ \rho _{23}  \ ,
\\ &  S  \ = \ S^2_{\triangle} \ = \ \frac{4\,\ta_2-\ta_1^2}{16} \ = \  \frac{ 2 (\rho _{12} \,\rho _{13}+
                      \rho _{12} \,\rho _{23}+ \rho _{13}\, \rho _{23}) - (\rho _{12}^2+\rho _{13}^2+\rho _{23}^2) }{16}\ ,
\\ & T=\ta_3 \ = \ \rho _{12} \, \rho _{13} \, \rho _{23} \ ,
\end{aligned}
\end{equation}
namely,
\begin{equation}
\label{addition3-3tauS}
\begin{aligned}
\De_R \ = &  \ 6\,P\,\pa^2_P + \frac{1}{2}\,P\,S\,\pa_{S}^2 +
   T\,(48\,S + P^2)\,\pa_{T}^2 + 36\,T\,\pa_{P,{T}} +
   24\,S\,\pa_{P,S} + 2\,S (16\,S + P^2)\,\pa_{S,{T}}
\\
  &\ +\ 6\,d\,\pa_P\ +\ \frac{1}{4}\,(d-1)\,P\,\pa_{S}\ +\ [8\,(d+4)\,S + \frac{d}{2}\,P^2]\,\pa_{T}  \ ,
\end{aligned}
\end{equation}
where the metric is of the form

\begin{equation}
\label{}
g^{\mu \nu}\ = \left|
  \begin{array}{ccc}
      6\,P & \ 12\,S & \ 18\,T \\
    12\,S & \ \frac{1}{2}P\,S & \ S(16\,S+P^2) \\
    18\,T & \ S(16\,S+P^2) & \ \tau_3(48\,S+P^2) \\
  \end{array}
\right|  \ ,
\end{equation}
with determinant
\[
\det g^{\mu \nu}\ = \   -3\, P\, S\, \left(\ 54\, T^2 \ - \ T\,P \left(P^2\,+\,144\, \,S\right)\ +\ 2 \,S\, \left(P^2+16\, S\right)^2\ \right)
\ .
\]
Note that the operator (\ref{addition3-3tauS}) is of Lie-algebraic nature: it can be rewritten in terms of the generators of the
algebra $h^{(3)}$ realized by the differential operators, see below. It acts as a filtration for the flag of finite-dimensional
representation spaces of this algebra
\[
     {\mathcal P}^{(1,2,3)}_{N}\ =\ \langle P^{p_1} S^{p_2} T^{p_3} \vert \
     0 \le p_1+2p_2+3p_3 \le N \rangle\ ,\ N=0,1,2, \ldots\ .
\]
This operator will be instrumental for  construction  of (quasi)-exactly-solvable problems for  Case III.

\subsection{Towards $d=1$}

For $d=1$ when the square of the interaction triangle vanishes, hence, $S \equiv \frac{1}{16}(4\ta_2-\ta_1^2)=0$, the metric becomes degenerate, $\det g^{\mu \nu}(\rho)=\det g^{\mu \nu}(\ta)=0$. Effectively, it leads to a reduction of the dimension of the space of relative distances from 3 to 2:
the configuration space $4\ta_2-\ta_1^2 \geq 0$ shrinks to the boundary $4\ta_2-\ta_1^2=0$. In order to see this dimensional reduction explicitly,
let us change variables in (\ref{addition3-3rho}),
\[
(\rho_{12}, \rho_{13}, \rho_{23}) \rar (\rho_{12}, \rho_{13}, S)\ .
\]
It follows that
\[
  {\De_R}(\rho)\ =\ 4(\rho_{12} \pa^2_{\rho_{12}} + \rho_{13} \pa^2_{\rho_{13}}) + \frac{1}{2} \,S\, (\rho_{12} + \rho_{13} + \rho_{23}) \pa^2_{S}\ +\ 2 (\rho_{12} + \rho_{13} - \rho_{23})\pa_{\rho_{12}}\pa_{\rho_{13}}\ +
\]
\begin{equation}
          8\, S\, (\pa_{\rho_{12}}\pa_{S}\ +\ \pa_{\rho_{13}}\pa_{S})\ +\ 2d (\pa_{\rho_{12}} + \pa_{\rho_{13}})+ \frac{1}{4} (d-1)(\rho_{12} + \rho_{13} + \rho_{23})  \pa_{S} \ ,
\label{addition3-3rho-A}
\end{equation}
with metric
\begin{equation}
\label{}
g^{\mu \nu}\ = \left|
  \begin{array}{ccc}
      4\,\rho_{12} & \ 4\,S & \ \rho_{12}+\rho_{13}-\rho_{23} \\
    4\,S & \ \frac{1}{2}\,S\,(\rho_{12}+\rho_{13}+\rho_{23}) & \ 4\,S \\
    \rho_{12}+\rho_{13}-\rho_{23} & \ 4\,S & \ 4\,\rho_{13} \\
  \end{array}
\right|  \ ,
\end{equation}
where ${\rho}_{23} = {\rho}_{23}({\rho}_{12}, {\rho}_{13}, S)$
and determinant
\[
\det g^{\mu \nu}\ = \ -12\,S\,(4\,S-\rho_{12}\,\rho_{13})\,{\rho}_{23}\ ,
\]
which vanishes identically if $S=0$.

Imposing conditions: $d=1$ and $S=0$ on $\De_R$ (\ref{addition3-3rho-A}), we get
\begin{equation}
  {{\tilde \De}_R}(\rho)\ =\ 4(\rho_{12} \pa^2_{\rho_{12}} + \rho_{13} \pa^2_{\rho_{13}})\ +\ 2 (\rho_{12} + \rho_{13} - \rho_{23})\pa_{\rho_{12}}\pa_{\rho_{13}}\
           +\ 2 (\pa_{\rho_{12}} + \pa_{\rho_{13}}) \ ,
\label{addition3-3rho-1}
\end{equation}
where ${\rho}_{23} =  (\sqrt{{\rho}_{12}} \pm \sqrt{{\rho}_{13}})^2$\,. The operator (\ref{addition3-3rho-1}) is not algebraic anymore.
However, it can be easily checked by calculating the curvature that is a Laplace-Beltrami operator in flat space.
By changing variables $\rho \rar r$ in (\ref{addition3-3rho-1}),
\[
r_{12}^2\ =\ \rho_{12}\ ,\ r_{13}^2\ =\ \rho_{13}\ ,
\]
we arrive at the algebraic operator (\ref{Drel3-1}), which is nothing but the flat Laplacian.
There are also numerous changes of the coordinates $\rho_{12}, \rho_{13}$, see Section II, into $\xi_{1,2}$ (\ref{d1-xi}),
$\si_{1,2}$ (\ref{d1-si}), $\la_{1,2}$ (\ref{d1-la}), $(t, u)$ {\it etc.} in all those the operator (\ref{addition3-3rho-1}) is algebraic.

It is important to emphasize that unlike (\ref{addition3-3rho-A}), in the limit $d\rightarrow 1$ and $S\rightarrow 0$ the
algebraic operator (\ref{addition3-3tauS}) continues to be algebraic,
\begin{equation}
\label{addition3-3tauS2}
\begin{aligned}
\De_R \ = &  \ 6\,P\,\pa^2_P \ + \
   T\, P^2\,\pa_{T}^2 \ + \  36\,T\,\pa_{P,{T}}
  \ +\ 6\,\pa_P\ +\ \frac{1}{2}\,P^2\,\pa_{T}  \ .
\end{aligned}
\end{equation}
The metric of operator (\ref{addition3-3tauS2}) is given by
\begin{equation}
\label{gmunu-d1}
g^{\mu \nu}\ = \  \left|
  \begin{array}{cc}
  \  6\,P & \ 18\,T \\
   \ 18\,T & \ T\,P^2 \\
  \end{array}
\right| \ ,
\end{equation}
with determinant
\[
\det g^{\mu \nu}\ = \  6\,T\,(P^3 - 54\,T)\ .
\]
Its curvature is zero, thus, $\De_R$ is the Laplace-Beltrami operator in  flat space. In fact, the geometrical
coordinates $P, T$ correspond to $\la_{1,2}$ (\ref{d1-la}), in those the 3-body $G_2$ rational, Wolfes model becomes algebraic.
Later the geometrical coordinates $P, S, T$ will play an important role in construction of (quasi)-exactly-solvable 3-body problems.

\subsection{Integral}

It can be shown that there exists the 1st order symmetry operator written in $\rho-$variables as
\begin{equation}
L_1 \ = \  (\rho_{13}-\rho_{23})\pa_{\rho_{12}} + (\rho_{23}-\rho_{12})\pa_{\rho_{13}} + (\rho_{12}-\rho_{13})\pa_{\rho_{23}} \ ,
\label{integral}
\end{equation}
for the operator (\ref{addition3-3rho}), such that
\[
[{\De_R}(\rho)\ ,\ L_1]=0\ .
\]
Here, $L_1$ is an algebraic operator, which is anti-invariant under the $S_3$-group action.
The existence of the symmetry $L_1$ implies that in the space of relative distances one variable
can be separated out in (\ref{addition3-3rho}).

Set
\begin{equation}
\label{wcoords1}
     w_1\ =\ \rho_{12}+\rho_{13}+\rho_{23}=P\ ,\quad
     w_2\ =\ 2\,
       \sqrt{\rho_{12}^2+\rho_{13}^2+\rho_{23}^2-\rho_{12}\rho_{13}-\rho_{12}\rho_{23} -\rho_{13}\rho_{23}}
       \quad ,
\end{equation}
where $w_2^2=4(\ta_1^2 - 3 \ta_2)=P^2-48S$ in geometrical variables as well\,, which are invariant under the action of $L_1$, and
\[
   w_3=\frac{\sqrt{3}}{9}\left({\rm sgn}\,(\rho_{23}-\rho_{13})\arcsin(\frac{2\rho_{12}-\rho_{23}-\rho_{13}}{w_2})
   +{\rm sgn}(\rho_{13}-\rho_{12})\,\arcsin(\frac{2\rho_{23}-\rho_{13}-\rho_{12}}{w_2})\right.
\]

\begin{equation}
\label{wcoords2}
    \left. + {\rm sgn}(\rho_{12}-\rho_{23})\,\arcsin(\frac{2\rho_{13}-\rho_{23}-\rho_{12}}{w_2})-\frac{3\pi}{4}\right),
\end{equation}
with ${\rm sgn}(x)=\frac{x}{|x|}$ for nonzero $x$. These coordinates are invariant under a cyclic permutation of the indices on the $\rho_{jk}$: $1\to 2\to 3\to 1$. Under a transposition of exactly two indices, see e.g. $(12), (3)$\,, we see that $w_1,w_2$ remain invariant,
and $w_3 \to -w_3-\frac{\sqrt{3}\pi}{6}$. (For the method used to compute $w_3$ see \cite{Ince}.)
Expressions for $w_3$ vary, depending on which of the 6 non-overlapping regions of
$(\rho_{12}, \rho_{13}, \rho_{23})$ space we choose to evaluate them:
\begin{enumerate}
 \item \[ (a):\ \rho_{23}>\rho_{13}>\rho_{12}\ ,\quad (b):\ \rho_{13}>\rho_{12}>\rho_{23}\ ,\quad (c):\ \rho_{12}>\rho_{23}>\rho_{13}\ ,\]
\item   \[(d):\ \rho_{13}>\rho_{23}>\rho_{12}\ ,\quad (e):\ \rho_{12}>\rho_{13}>\rho_{23}\ ,\quad (f):\ \rho_{23}>\rho_{12}>\rho_{13}\ ,\]
\end{enumerate}

The regions in class 1 are related by cyclic permutations, as are the regions in class 2.
We  map between regions by a transposition. Thus it is enough to evaluate $w_3$ in the region $(a):\ \rho_{23}>\rho_{13}>\rho_{12}$. The other 5 expressions will then follow from the
permutation symmetries. In this case we have
\[ (a):\  w_3=-\frac{\sqrt{3}}{9}\arcsin\left[\frac{2\sqrt{2}}{w_2^3}((2-\sqrt{3})\rho_{13}-\rho_{23}+(\sqrt{3}-1)\rho_{12})\times\right.\]
\[\left.(2\rho_{23}
 -(1+\sqrt{3})\rho_{13}+(\sqrt{3}-1)\rho_{12})((2+\sqrt{3})\rho_{12}-(1+\sqrt{3})\rho_{13}-\rho_{23})\right].\]
 (The special cases where exactly two of the $\rho_{jk}$ are equal can be obtained from these results by continuity. Here, $w_3$ is a single-valued differentiable function of
 $\rho_{12},\rho_{13},\rho_{23}$ everywhere in the physical domain (configuration space), except for the points $\rho_{12}=\rho_{13}=\rho_{23}$ where it is undefined.)

In  these coordinates, the operators (\ref{integral}) and  (\ref{addition3-3rho}) take the form
\[
 L_1(w) \ = \pa_{w_3}\ ,
\]
\[
 \frac{1}{6}\De_R(w) \ =\ \ w_1\pa_{w_1}^2\ +\ w_1\pa_{w_2}^2\ +\ \frac{w_1}{3w_2^2}\pa_{w_3}^2
 \ +\ 2\,w_2\pa_{w_1w_2}^2\ +\ d\,\pa_{w_1}\ +\ \frac{w_1}{w_2}\pa_{w_2} \ ,
\]
respectively. It is evident that for the $w_3$-independent potential
\[
    V(w_1,\,w_2; w_3) \ = \  6\,g(w_1,\,w_2) \ ,
\]
where the factor 6 is introduced for convenience,
the operator $L_1$ is still an integral for an arbitrary function $g$:
\[
[L_1(w) , -\De_R(w) + 6\,g(w_1,\,w_2)] = 0\ .
\]
This property of integrability permits  separation of the variable $w_3$ in the spectral
problem
\[
   [-\De_R(w) + 6\,g(w_1,\,w_2)]\Psi \ =\ E \Psi\ ,
\]
where $\Psi = \psi(w_1,\,w_2)\,\xi(w_3)$ is defined by the differential equations,
\begin{equation}
\label{eq-xi}
   \pa_{w_3} \xi\ =\ i\,p\, \xi\ ,\quad \xi\,=\,e^{i p w_3}\ ,\quad p=0,\pm 1,\pm 2\ \ldots
\end{equation}
\[
  \left(w_1\pa_{w_1}^2\ +\ w_1\pa_{w_2}^2
 \ +\ 2\,w_2\pa_{w_1w_2}^2\ +\ d\,\pa_{w_1}\ +\ \frac{w_1}{w_2}\pa_{w_2}
  \ -\ p^2\,\frac{w_1}{3\,w_2^2}\ -\ g(w_1,\,w_2) \right) \psi\
\]
\begin{equation}
\label{eq-psi}
  =\ -E\, \psi\ .
\end{equation}

Note that the integral $L_1$ is the integral for the three-dimensional quantum problem (\ref{Hrel-final}). As for the original
3-body problem (\ref{Hrel}) this integral is a {\it particular} integral \cite{Turbiner:2013p}: it commutes with the Hamiltonian (\ref{Hrel})
over the space of relative distances ${\bf \tilde R}$ only
\[
   [{\cal H}_r , L_1]\, :\, {\bf \tilde R}\ \rar \ \{0\}\ .
\]
As for the operators ${\cal H}_r$ and $L_1$, they do not commute.

\subsection{The Representations of $sl(4,{\bf R})$}

Both operators (\ref{addition3-3rho}) and (\ref{integral}) are $sl(4,{\bf R})$-Lie algebraic - they can be rewritten in terms of the
generators of the maximal affine subalgebra $b_4$ of the algebra $sl(4,{\bf R})$, see e.g. \cite{Turbiner:1988,Turbiner:2016}
\begin{eqnarray}
\label{sl4R}
 {\cal J}_i^- &=& \frac{\pa}{\pa u_i}\ ,\qquad \quad i=1,2,3\ , \non  \\
 {{\cal J}_{ij}}^0 &=&
               u_i \frac{\pa}{\pa u_j}\ , \qquad i,j=1,2,3 \ , \\
 {\cal J}^0(N) &=& \sum_{i=1}^{3} u_i \frac{\pa}{\pa u_i}-N\, , \non \\
 {\cal J}_i^+(N) &=& u_i {\cal J}^0(N)\ =\
    u_i\, \left( \sum_{j=1}^{3} u_j\frac{\pa}{\pa u_j}-N \right)\ ,
       \quad i=1,2,3\ ,
\end{eqnarray}
where $N$ is a parameter and
\[
 u_1\equiv\rho_{12}\ ,\qquad u_2\equiv\rho_{13}\ , \qquad u_3\equiv\rho_{23} \ .
\]
If $N$ is a non-negative integer, a finite-dimensional representation space occurs,
\begin{equation}
\label{P3}
     {\cal P}_{N}\ =\ \langle u_1^{p_1} u_2^{p_2} u_3^{p_3} \vert \ 0 \le p_1+p_2+p_3 \le N \rangle\ .
\end{equation}
It is easy to check that the space ${\cal P}_{N}$ is invariant with respect to projective transformations,
\[
   u_i \rar \frac{a_i u_1 + b_i u_2 + c_i u_3 + d_i}{\al u_1 + \beta u_2 + \gamma u_3 + \delta}\ ,\ i=1,2,3
   \ ,
\]
where $a_i, b_i, c_i, d_i, \al , \beta, \gamma, \delta$ are real parameters, taking them as the rows of the 4 x 4 matrix $G$ we arrive at the
condition $G \in GL(4,R)$.

Explicitly, above-mentioned operators $\De_R, L_1$ take the form
\begin{equation}
\label{HRex}
  \frac{1}{2}\, \De_R({\cal J}) \ = \ 2(\, {\cal J}_{11}^0\,{\cal J}_1^- + {\cal J}_{22}^0\,{\cal J}_2^-
      + {\cal J}_{33}^0\,{\cal J}_3^-  \,)\
      +\ d \,({\cal J}_1^- + {\cal J}_2^- + {\cal J}_3^-)\ +
\end{equation}
\[
      \bigg({\cal J}_{11}^0\,({\cal J}_2^- + {\cal J}_3^- ) +  {\cal J}_{22}^0\,({\cal J}_1^- +
      {\cal J}_3^-  ) +   {\cal J}_{33}^0\,({\cal J}_1^- + {\cal J}_2^-  ) -
      {\cal J}_{31}^0\,{\cal J}_2^- - {\cal J}_{23}^0\,{\cal J}_1^- - {\cal J}_{12}^0\,{\cal J}_3^- \bigg)
      \ ,
\]
and
\begin{equation}
     L_1 \ = \  {\cal J}_{21}^0\,-\,{\cal J}_{31}^0\, + \,{\cal J}_{32}^0\,-{\cal J}_{12}^0\, +
     \, {\cal J}_{13}^0\,-\,{\cal J}_{23}^0\, \ .
\label{integral-J}
\end{equation}
in terms of $sl(4,{\bf R})$ generators.

\subsection{The Laplace-Beltrami operator, underlying geometry}

A remarkable property of the algebraic operator ${\De_R}(\rho)$ (\ref{addition3-3rho}) is its gauge-equivalence to the Schr\"odinger operator.
Making the gauge transformation with determinant $D(\rho)$\ (\ref{gmn33-rho-det}), (\ref{gmn33-rho-det-gamma}) included into the factor for $d \neq 1$,

\[
\Gamma \ = \ D^{-\frac{1}{4}}(4 \ta_2-\ta_1^2)^{\frac{3-d}{4}}  \ \sim \ \frac{1}{{\ta_1^{\frac{1}{4}}(4 \ta_2-\ta_1^2)}^{\frac{d-2}{4}}}  \ \sim \ P^{-\frac{1}{4}}\ S^{\frac{2-d}{2}}_{\triangle} \ ,
\]
see also (\ref{taus}), we find that
\begin{equation}
         \Gamma^{-1}\, {\De_R}(\rho)\,\Gamma \ =
        \  \De_{LB}(\rho) - \tilde V(\rho) \ ,
\label{HLB3}
\end{equation}
where the effective potential is of the form
\[
  \tilde V(\rho)\ = \ \frac{9}{8 \left(\rho_{12}+\rho_{13}+\rho_{23}\right)}\ -\
\frac{(d-2)(d-4)}{2}  \,
  \frac{\left(\rho_{12}+\rho_{13}+\rho_{23}\right)}
  {\left(\rho_{12}^2+\rho_{13}^2+\rho_{23}^2 -2 \rho_{12} \rho_{13}-
                     2 \rho_{12} \rho_{23}-2 \rho_{13} \rho_{23}\right)}
\]

\begin{equation}
   =\ \frac{9}{8\,P}\ +\ \frac{(d-2)(d-4)}{32} \frac{P}{S^2_{\triangle}}  \ ,
\label{Veff}
\end{equation}
thus, of geometric nature, with vanishing second term at $d=2, 4$.
In turn,
\[
  \De_{LB}(\rho) \ =\ 4(\rho_{12} \pa^2_{\rho_{12}} + \rho_{13} \pa^2_{\rho_{13}} +\rho_{23} \pa^2_{\rho_{23}})
\]
\[
   + \ 2\, \bigg((\rho_{12} + \rho_{13} - \rho_{23})\pa_{\rho_{12}}\pa_{\rho_{13}}\ +
          (\rho_{12} + \rho_{23} - \rho_{13})\pa_{\rho_{12}}\pa_{\rho_{23}}\ +
          (\rho_{13} + \rho_{23} - \rho_{12})\pa_{\rho_{13}}\pa_{\rho_{23}}
  \bigg)
\]
\begin{equation}
\label{LB3}
- 3\, \bigg(\frac{\rho_{12}\pa_{\rho_{12}}+\rho_{13}\pa_{\rho_{13}}+\rho_{23}\pa_{\rho_{23}}}
{\rho _{12}+\rho _{13}+\rho _{23}} \bigg) + 4\, (\pa_{\rho_{12}}+\pa_{\rho_{23}}+ \pa_{\rho_{13}})\ ,
\end{equation}
is the $d$-independent Laplace-Beltrami operator,
\[
   \De_{LB}(\rho)\ =\ {\sqrt {D(\rho)}}\, \pa_{\mu} \frac{1}{\sqrt {D(\rho)}}\, g^{\mu \nu} \pa_{\nu}\ ,\quad \pa_{\nu}\equiv \frac{\pa}{\pa{\rho_{\nu}}}\ ,
\]
see (\ref{gmn33-rho}), (\ref{gmn33-rho-det}). Eventually, taking into account the gauge rotation (\ref{HLB3}) we arrive at the three-dimensional Hamiltonian
\begin{equation*}
\label{H-3-3r-r}
    {H}_{LB} (r) \ =\ -\De_{LB}(r) + \tilde V (r) +  V(r)
    \ ,
\end{equation*}
in the space of $r$-relative distances, or
\begin{equation*}
\label{H-3-3r-rho}
    {H}_{LB} (\rho) \ =\ -\De_{LB}(\rho) + \tilde V (\rho) +  V(\rho)\ ,
\end{equation*}
in $\rho$-space, see (\ref{rho}), or
\begin{equation*}
\label{H-3-3r-tau}
    {H}_{LB} (\tau) \ =\ -\De_{LB}(\tau) + \tilde V (\tau) +  V(\tau)\ ,
\end{equation*}
in $\tau$-space, see (\ref{taus}). These Hamiltonians
describe the three-dimensional quantum particle moving in the curved space with metric $g^{\mu \nu}$ with kinetic energy $\De_{LB}$, in particular, in $\rho$-space with metric $g^{\mu \nu}(\rho)$ (\ref{gmn33-rho}) with kinetic energy $\De_{LB}(\rho)$ (\ref{LB3}) and effective potential $\tilde V(\rho)$ (\ref{Veff}). The Ricci scalar, see e.g. \cite{Eis}, in $\rho$-space is equal to
\[
  Rs \ = \ -\frac{41\,{(\rho_{12}+\rho_{13}+\rho_{23})}^2  -84\,(\rho_{12}\,\rho_{13}+\rho_{12}\,\rho_{23}+\rho_{23}\,\rho_{13})}{12\,(\rho_{12}+\rho_{13}+\rho_{23})
   \left({(\rho_{12}+\rho_{13}+\rho_{23})}^2-4\,(\rho_{12}\,\rho_{13}+\rho_{12}\,\rho_{23}
   +\rho_{23}\,\rho_{13})\right)}
\]
\[
\ = \ \frac{-84\,\tau_2\ +\ 41\,{\tau_1}^2   }{12\,\tau_1(4\,\tau_2\ -\ \tau_1^2)}
\ =\
     \frac{5 P^2 - 84 S}{48\,P\, S}\ =\ -\frac{7}{4\,P} + \frac{5\,P}{48\,S}\ ,
\]
interestingly, it has a structure similar to one of the effective potential (\ref{HLB3}).

It is singular at the boundary of the configuration space. (For $d=1$ the configuration space degenerates to the
boundary, $4\,\tau_2\ =\ \tau_1^2$ and becomes flat.) The Cotton tensor, see e.g. \cite{Eis}, for the metric (\ref{gmn33-rho}) is nonzero,
so the space is {\it not} conformally flat.

Making the de-quantization of (\ref{H-3-3r-rho}) we arrive at a three-dimensional classical system which is characterized by the Hamiltonian,
\begin{equation}
\label{H-3-3r-rho-class}
    {H}_{LB}^{(c)} (\rho) \ =\ g^{\mu \nu}(\rho)\,P_{\mu}\, P_{\nu} \ + \  \tilde V (\rho) \ + \   V(\rho)\ ,
\end{equation}
where $P_{\mu}\,, \ \mu=12,23,13$ is classical canonical momenta in $\rho$-space and $g^{\mu \nu}(\rho)$ is given by (\ref{gmn33-rho}). Here the underlying manifold (zero-potential case) admits an $so(3)$ algebra
of constants of the motion linear in the momenta, i.e., Killing vectors. Thus, the free Hamilton-Jacobi equation is integrable. However,
it admits no separable coordinate system. The classical kinetic energy
$T=g^{\mu \nu}(\rho)\,P_{\mu}\, P_{\nu}$  Poisson-commutes with
\[
 L_1^{(c)} \ = \  (\rho_{13}-\rho_{23})P_{12} + (\rho_{23}-\rho_{12})P_{13} + (\rho_{12}-\rho_{13})P_{23}\ .
\]

\section{{(Quasi)}-exact-solvability}

\subsection{QES in $\rho, \ta-$ variables, $d \neq 1$}

{\bf (I).\ \it Quasi-Exactly-Solvable problem in $\rho$-variables.}

Let us take the $d$-independent function
\begin{equation}
\label{psi_cal-r-d3}
       \Psi_0(\rho_{12},\,\rho _{13},\,\rho _{23}) \ = \ \ta_1^{1/4}
       {(4\,\ta_2-\tau_1^2)}^{\frac{\gamma}{2}}\,e^{-\om\,\ta_1 - \frac{A}{2}\,\ta_1^2} \
       \equiv \Psi_0(\ta_1,\ta_2)\ ,
\end{equation}
assuming $d \neq 1$, where $\gamma,\,\om > 0$ and $A \geq 0$ and for $\om=0$, $A>0$ are constants and $\ta$'s are given by (\ref{taus}). We look for the potential for which the function (\ref{psi_cal-r-d3}) is the ground state function for the Hamiltonian ${H}_{LB}(\rho)$ of the 3-dimensional quantum particle. This potential can be found immediately by calculating the ratio
\[
\frac{\De_{LB}(\rho) \Psi_0 }{ \Psi_0}\ =\ V_0 - E_0 \ ,
\]
where $\De_{LB}(\rho)$ is given by (\ref{LB3}) with metric (\ref{gmn33-rho}).
The  result is
\[
  V_0(\ta_1,\,\ta_2)\ = \ \frac{9}{8 \ta_1}  + \,\gamma(\gamma-1)
  \left ( \frac{2\ta_1}{4\ta_2-\tau_1^2} \right )\ +
\]
\begin{equation}
 6\,\om^2\,\ta_1\, +\, 6\,A\,\ta_1\,(2\,\om\,\ta_1\, -\, 2\gamma - 3)\,
   +\, 6\, A^2\ta_1^3      \ ,
\label{VQES-0}
\end{equation}
which is $d$-independent, it includes the effective potential $\tilde V$ and many-body potential $V$ with the energy of the ground state
\begin{equation}
  E_0\ =\ 12\,\om\,(1+\,\gamma) \ .
\label{EQES-0}
\end{equation}

Now, let us take the Hamiltonian ${H}_{LB,0} \equiv  -\De_{LB}(\rho) + V_0$ with potential (\ref{VQES-0}), subtract $E_0$ (\ref{EQES-0}) and make the gauge
rotation with $\Psi_0$ (\ref{psi_cal-r-d3}). As the result we obtain the $sl(4, {\bf R})$-Lie-algebraic operator with additional potential $\De V_N$, \cite{Turbiner:1988,Turbiner:2016}
\[
   \Psi_0^{-1}\,(-{\De_{LB}}(\rho) + V_0 - E_0)\,\Psi_0\ =\ -{\De_R}({\cal J})\
   +\ 2(d - 2 - 2\,\gamma)\,({\cal J}_1^- + {\cal J}_2^- + {\cal J}_3^-)\ +
\]
\begin{equation}
     12\,\om\,({\cal J}_{11}^0  +{\cal J}_{22}^0 + {\cal J}_{33}^0) + 12\,A\,\left({\cal J}_1^+(N) + {\cal J}_2^+(N)  + {\cal J}_3^+(N) \right)\ +\ 12 \,A\,N \ta_1
\label{HQES-0-Lie}
\end{equation}
\[
  \equiv h^{(qes)}(J)\ +\ \De V_N\ ,
\]
see (\ref{HRex}), where
\[
    \De V_N\ =\ 12 \,A\,N \ta_1\ .
\]
It is evident that for integer $N$ the $d$-independent operator $h^{(qes)}(J)$ has a finite-dimensional invariant subspace ${\cal P}_{N}$, (\ref{P3}), with $\dim {\cal P}_{N} \sim N^3$ at large $N$.
Finally, we arrive at the quasi-exactly-solvable, $d$-independent, single particle Hamiltonian in the space of relative distances $\rho$,
\begin{equation}
\label{HQES-3-3}
    {H}_{LB}^{(qes)}(\rho) \ =\ -\De_{LB}(\rho) \ + \  V_N^{(qes)}(\rho)\ ,
\end{equation}
cf.(\ref{Hrel-final}), where
\[
   V^{(qes)}_N(\ta_1,\,\ta_2)\ =      \ \frac{9}{8 \ta_1}  + \,\gamma(\gamma-1)\left ( \frac{2\ta_1}{4\ta_2-\ta_1^2}\right )\ +
\]
\begin{equation}
  +\ 6\,\om^2\,\ta_1\
  +\ 6\,A\,\ta_1\,(2\,\om\,\ta_1\,-\,2\gamma\ -\ 2 N\, -\, 3)\ +\ 6\, A^2\ta_1^3 \ .
\label{VQES-N}
\end{equation}
is two-variable QES potential. Its configuration space is $\ta_1 \geq 0$ and $4\ta_2 \geq \ta_1^2$.
For this potential $\sim N^3$ eigenstates can be found by algebraic means. They have the factorized form of the polynomial in $\rho$ multiplied by $\Psi_0$ (\ref{psi_cal-r-d3}),
\[
   \mbox{Pol}_N (\rho_{12}, \rho_{13},\rho_{23})\ \Psi_0 (\ta_1, \ta_2)\ .
\]
These polynomials are the eigenfunctions of the quasi-exactly-solvable algebraic operator
\begin{equation}
\label{hQES-N-alg}
  \frac{1}{2}\,h^{(qes)}(\rho) \ = \
\end{equation}
\[
   -2(\rho_{12} \pa^2_{\rho_{12}} + \rho_{13} \pa^2_{\rho_{13}} +\rho_{23} \pa^2_{\rho_{23}})
\]
\[
 - \bigg( (\rho_{12} + \rho_{13} - \rho_{23})\pa_{\rho_{12}}\pa_{\rho_{13}}
          +(\rho_{12} + \rho_{23} - \rho_{13})\pa_{\rho_{12}}\pa_{\rho_{23}}
          +(\rho_{13} + \rho_{23} - \rho_{12})\pa_{\rho_{13}}\pa_{\rho_{23}}\bigg)
\]
\[
    -\ 2\,(1\,+\,\gamma)(\pa_{\rho_{12}} + \pa_{\rho_{13}}+ \pa_{\rho_{23}}) + 6\,\om(\rho_{12}\pa_{\rho_{12}}+\rho_{13}\pa_{\rho_{13}}+\rho_{23}\pa_{\rho_{23}})
\]
\[
   - 6\,A\,(\rho_{12}+ \rho_{13}+\rho_{23})(\rho_{12}\,\pa_{\rho_{12}} + \rho_{13}\,\pa_{\rho_{13}}+ \rho_{23}\,\pa_{\rho_{23}}-N   )
\]
which is the quasi-exactly-solvable $sl(4,\,{\bf R})$-Lie-algebraic operator
\begin{equation}
   \frac{1}{2}\,h^{(qes)}(J) \ = \  -2(\, {\cal J}_{11}^0\,{\cal J}_1^- + {\cal J}_{22}^0\,{\cal J}_2^-
      + {\cal J}_{33}^0\,{\cal J}_3^-  \,)
\label{hQES-N-Lie}
\end{equation}
\[
 - \bigg({\cal J}_{11}^0\,({\cal J}_2^- + {\cal J}_3^- ) +  {\cal J}_{22}^0\,({\cal J}_1^- +
      {\cal J}_3^-  ) +   {\cal J}_{33}^0\,({\cal J}_1^- + {\cal J}_2^-  ) -
      {\cal J}_{31}^0\,{\cal J}_2^- - {\cal J}_{23}^0\,{\cal J}_1^- - {\cal J}_{12}^0\,{\cal J}_3^- \bigg)
\]
\[
   -\ 2\,(1\, +\,\gamma)\,( {\cal J}_1^- + {\cal J}_2^- + {\cal J}_3^-  ) + 6\,\om\,({\cal J}_{11}^0 + {\cal J}_{22}^0 + {\cal J}_{33}^0)
\]
\[
    +\, 6 \,A\,(\, J_1^+(N)+J_2^+(N)+J_3^+(N)\, )  \ ,
\]
cf. (\ref{HQES-0-Lie}).

As for the original many-body problem (\ref{Hrel-Mod}) in the space of relative motion
\[
   {\tilde {\cal H}}_R\,\Psi(r) \equiv \ \bigg(- {\De_R}(r)\ + \ V(r)\bigg)\, \Psi(r)\ =\ E\,\Psi(r)\ ,\ \Psi \in L_2 ({\bf \tilde R})\ ,
\]
the potential for which quasi-exactly-solvable, polynomial solutions occur in the form
\[
   \mbox{Pol}_N (\rho_{12}, \rho_{13},\rho_{23})\ \Gamma \  \Psi_0 (\ta_1, \ta_2) \ ,
\]
where $\Gamma \sim   D^{-1/4}(\rho)\,(4 \ta_2-\ta_1^2)^{\frac{3-d}{4}}$, see (\ref{gmn33-rho-det-gamma}).
The three-body potential is given by
\[
V_{relative, N}^{(qes)}(\ta) \ = \   V^{(qes)}_N \ - \ \tilde{V}\ =
\]
\begin{equation}
 \frac{4\,\gamma(\gamma-1)-(d-2)(d-4)}{2}\left (\frac{\ta_1}{4\ta_2-\ta_1^2}\right ) +\ 6\,\om^2\,\ta_1\
  +\ 6\,A\,\ta_1\,(2\,\om\,\ta_1\,-\,2\,{\gamma}\ -\ 2 N\, -\, 3)\
   +\ 6\, A^2\ta_1^3      \ ,
\label{VQES-N-rel}
\end{equation}
c.f. (\ref{VQES-N}); it does not depend on the $\ta_3$-variable and does not contain a singular term $\sim 1/\ta_1$.

\bigskip

{\bf (II).\ \it Exactly-Solvable problem in $\rho$-variables.}

If the parameter $A$ vanishes in (\ref{psi_cal-r-d3}), (\ref{VQES-N}) and (\ref{HQES-0-Lie}), (\ref{hQES-N-Lie}) we will arrive at the
exactly-solvable problem, where $\Psi_0$ (\ref{psi_cal-r-d3}) at $A=0$ plays the role of the ground state function,
\begin{equation}
\label{psi_cal-r-d2exact}
   \Psi_0(\rho_{12},\,\rho _{13},\,\rho _{23}) \ = \ \ta_1^{1/4}
    {(4\,\ta_2-\ta_1^2)}^{\frac{\gamma}{2}}\,e^{-\om\,\ta_1} \ .
\end{equation}
The $sl(4, {\bf R})$-Lie-algebraic operator (\ref{hQES-N-Lie}) contains no raising generators $\{{\cal J}^+(N)\}$ and becomes
\[
  h^{(exact)} = -{\De_R}({\cal J}) + 2(d - 2 -2\,\gamma)\,({\cal J}_1^- + {\cal J}_2^- + {\cal J}_3^-) +
     12\,\om\,({\cal J}_{11}^0  +{\cal J}_{22}^0 + {\cal J}_{33}^0)\ ,
\]
see (\ref{HRex}), and, hence, preserves the infinite flag of finite-dimensional invariant subspaces
${\cal P}_{N}$ (\ref{P3}) at $N=0,1,2\ldots$\,. The single particle potential (\ref{VQES-N}) becomes
\begin{equation}
   V^{(es)}(\ta_1,\,\ta_2)\ = \ \frac{9}{8 \ta_1} + \,\gamma(\gamma-1)\left ( \frac{2\ta_1}{4\ta_2-\ta_1^2}\right )\ +
  \ 6\,\om^2\,\ta_1\ =\
\label{VES}
\end{equation}
\[
   =\  \frac{9}{8 \left(\rho _{12}+\rho _{13}+\rho _{23}\right)}\ +\ 6\om^2 \left(\rho _{12}+\rho _{13}+
       \rho _{23}\right)
\]
\[
-\ \gamma(\gamma-1)
\left ( \frac{2(\rho_{12}+\rho_{13}+\rho_{13})}{\rho_{12}^2+\rho_{13}^2+\rho_{23}^2-2\rho_{12}\rho_{13}-2\rho_{12}\rho_{23}
-2\rho_{13}\rho_{23}}
\right )\
\]

%

Eventually, we arrive at the exactly-solvable single particle Hamiltonian in the space of relative distances
\begin{equation}
\label{HES-3-2}
    {H}_{LB}^{(es)}(\rho) \ =\ -\De_{LB}(\rho) \ + \  V^{(es)}(\rho)\ ,
\end{equation}
where the spectra of energies
\[
    E_{N=n_1+n_2+n_3}\ =\  12\, \om\, (N + \gamma + 1)\ ,\quad N=0,1,2,\ldots
\]
is equidistant. Its degeneracy is equal to number of partitions of $N=n_1 + n_2 + n_3$.
All eigenfunctions have the factorized form of a polynomial in $\rho$ multiplied by $\Psi_0$ (\ref{psi_cal-r-d2exact}),
\[
   \mbox{Pol}_N (\rho_{12}, \rho_{13},\rho_{23})\ \Psi_0 (\ta_1, \ta_2)\ ,\quad N=0,1,\ldots\ .
\]

These polynomials are eigenfunctions of the exactly-solvable, $d$-independent, algebraic operator
\[
    \frac{1}{2}\,h^{(exact)}(\rho)\ =
\]
\[
    \ - 2 (\rho_{12} \pa^2_{\rho_{12}} + \rho_{13} \pa^2_{\rho_{13}} +\rho_{23} \pa^2_{\rho_{23}})
    -\ 2\,(1\,+\,\gamma)(\pa_{\rho_{12}} + \pa_{\rho_{13}}+ \pa_{\rho_{23}}) + 6\,\om(\rho_{12}\pa_{\rho_{12}}+\rho_{13}\pa_{\rho_{13}}+\rho_{23}\pa_{\rho_{23}})
\]
\begin{equation}
\label{hES-rho}
          -\, (\rho_{12} + \rho_{13} - \rho_{23})\pa_{\rho_{12}}\pa_{\rho_{13}}
          -\, (\rho_{12} + \rho_{23} - \rho_{13})\pa_{\rho_{12}}\pa_{\rho_{23}}
          -\, (\rho_{13} + \rho_{23} - \rho_{12})\pa_{\rho_{13}}\pa_{\rho_{23}} \ ,
\end{equation}
or, equivalently, of the exactly-solvable $sl(4,{\bf R})$-Lie-algebraic operator
\[
     \frac{1}{2}\,h^{(exact)}(J) \ = \  -2(\, {\cal J}_{11}^0\,{\cal J}_1^- + {\cal J}_{22}^0\,{\cal J}_2^-
      + {\cal J}_{33}^0\,{\cal J}_3^-  \,)
\]
\[
  - \,\bigg({\cal J}_{11}^0\,({\cal J}_2^- + {\cal J}_3^- ) +  {\cal J}_{22}^0\,({\cal J}_1^- +
      {\cal J}_3^-  ) +   {\cal J}_{33}^0\,({\cal J}_1^- + {\cal J}_2^-  ) -
      {\cal J}_{31}^0\,{\cal J}_2^- - {\cal J}_{23}^0\,{\cal J}_1^- - {\cal J}_{12}^0\,{\cal J}_3^- \bigg)
\]
\begin{equation}
\label{hES-N-Lie}
    -\ 2\,(1\,+\,\gamma)\,( {\cal J}_1^- + {\cal J}_2^- + {\cal J}_3^-  ) + 6\,\om\,({\cal J}_{11}^0 +
    {\cal J}_{22}^0 + {\cal J}_{33}^0)\ .
\end{equation}
Those polynomials $\mbox{Pol}_N$ are orthogonal w.r.t. $\Psi_0^2$\,,  (\ref{psi_cal-r-d3}) at $A=0$, their domain is given by (\ref{CFrho}).
Being written in variables $w_{1,2,3}$, see above, they are factorisable, $F(w_1,w_2)\, f(w_3)$.
To the best of our knowledge these orthogonal polynomials have not been studied in literature.

The Hamiltonian with potential (\ref{VES}) can be considered as a type of a $d$-dimensional generalization of the 3-body Calogero model \cite{Calogero:1969}, see also \cite{RT:1995}, \cite{ST:2015}, with loss of the property of pairwise interaction and absence of singular interaction terms $\sim \frac{\tau_2}{\tau_3}$. Now the potential of interaction contains two- and three-body interaction terms. If $\gamma=0,1$ in (\ref{VES})
we arrive at the celebrated harmonic oscillator potential in the space of relative distances, see e.g. \cite{Green}. In turn, in the space of relative motion this potential contains no singular terms at all and becomes,
\[
     V \ =\ 6\,\om^2\, \ta_1\ =\ 6\,\om^2\,(\rho_{12} + \rho_{13} + \rho_{23})\ =\ 6\,\om^2\,(r^2_{12} + \ r^2_{13} + \ r^2_{23})\ ,
\]
see \cite{Green}. We arrive at the harmonic oscillator potential $V$. Therefore,
the potential (\ref{VES}) is a $d$-dimensional generalization of the harmonic oscillator in the space of relative motion rather than 3-body (rational) Calogero model. An attempt to construct a true $d$-dimensional  generalization of the 3-body (rational) Calogero model will be made below.

\bigskip

{\bf (III).\ \it (Quasi)-Exactly-Solvable problem in $\ta$-variables.}

The quasi-exactly-solvable $sl(4,\,{\bf R})$-Lie-algebraic operator $h^{(qes)}(J)$\,, (\ref{hQES-N-Lie}) as well as the exactly-solvable operator $h^{(es)}(J)$ (\ref{hES-N-Lie})  as its degeneration at $A=0$,
were acting originally in $\rho$ variables (\ref{hQES-N-alg}).  They are characterized by {\it accidental} permutation symmetry $S^3$ in
$\rho$ variables and, hence, they can be rewritten in $\ta$ variables (\ref{taus}). Surprisingly, the operator (\ref{hQES-N-alg}) remains algebraic (!)
\begin{equation}
\label{hQES-N-tau}
    h^{(qes)}(\tau) \ = \ -6\,\ta_1\pa_{\ta_1}^2 -2\ta_1(7\ta_2-\ta_1^2)\pa_{\ta_2}^2 -2\ta_3(6\ta_2-\ta_1^2)\pa_{\ta_3}^2
   -\,24\,\ta_2\pa_{\ta_1,\ta_2}^2 - 36\ta_3\pa_{\ta_3,\ta_3}^2\ -
\end{equation}
\[
2\,(4\ta_2^2+9\ta_1\ta_3-\ta_1^2\ta_2)\pa_{\ta_2,\ta_3}^2 -18\pa_1 -14\ta_1\pa_2-2(7\ta_2-\ta_1^2)\pa_{\ta_3}\
\]
\[
 -\ 4\,(1\,+\,\gamma)\,(3\pa_{\ta_1}+2\ta_1\pa_{\ta_2}+\ta_2\pa_{\ta_3}) +12\om\,(\ta_1\pa_1+2\ta_2\pa_2+3\ta_3\pa_3)\ +
\]
\[
  12\,A\ta_1(\ta_1\pa_{\ta_1} + 2\ta_2\pa_{\ta_2} + 3\ta_3\pa_{\ta_3} - N) \ .
\]
Evidently, it remains algebraic at $A=0$ as well,
\begin{equation}
\label{hES-N-tau}
    h^{(es)}(\tau) \ = \ -6\,\ta_1\pa_{\ta_1}^2 -2\ta_1(7\ta_2-\ta_1^2)\pa_{\ta_2}^2 -2\ta_3(6\ta_2-\ta_1^2)\pa_{\ta_3}^2
   -\,24\,\ta_2\pa_{\ta_1,\ta_2}^2 - 36\ta_3\pa_{\ta_3,\ta_3}^2\ -
\end{equation}
\[
 2\,(4\ta_2^2+9\ta_1\ta_3-\ta_1^2\ta_2)\pa_{\ta_2,\ta_3}^2 -18\pa_1 -14\ta_1\pa_2-2(7\ta_2-\ta_1^2)\pa_{\ta_3}\
\]
\[
 -\ 4\,(1\,+\,\gamma)\,(3\pa_{\ta_1}+2\ta_1\pa_{\ta_2}+\ta_2\pa_{\ta_3}) +12\om\,(\ta_1\pa_{\ta_1}+2\ta_2\pa_{\ta_2}+3\ta_3\pa_{\ta_3})\ ,
\]
becoming the exactly-solvable one. Note that both quasi-exactly-solvable operator (\ref{hQES-N-tau}) and exactly-solvable operator (\ref{hES-N-tau}) admit the integral
\begin{equation}
\label{integral-tau}
     -L_1^2 \ =\
     \left(27\tau_3^2 - 18\tau_3\tau_2\tau_1 + 4\tau_3\tau_1^3 + 4\tau_2^3 - \ta_2^2 \ta_1^2  \right) \pa^2_{\tau_3}\ +\ \left(27\tau_3 - 9\tau_1\tau_2 + 2\tau_1^3 \right) \pa_{\tau_3}\ ,
\end{equation}
cf. (\ref{integral}), $[h^{(qes)}(\tau), L_1^2]=0$. This integral is an algebraic operator. It involves derivatives w.r.t. $\ta_3$ only.

It can be immediately checked that the quasi-exactly-solvable operator (\ref{hQES-N-tau}) has the finite-dimensional invariant subspace in polynomials in $\ta$,
\begin{equation}
\label{P3-tau}
     {\mathcal P}^{(1,2,3)}_{N}\ =\ \langle \ta_1^{p_1} \ta_2^{p_2} \ta_3^{p_3} \vert \
     0 \le p_1+2p_2+3p_3 \le N \rangle\ ,
\end{equation}
cf. (\ref{P3}), with characteristic vector $(1,2,3)$, hence, the Newton pyramid has sides 1,2,3, associated with solid angle of $90^o$,
for discussion see e.g. \cite{Turbiner:2013}. This finite-dimensional space appears as a finite-dimensional representation space of the algebra
of differential operators $h^{(3)}$ which was discovered in the relation with $H_3$ (non-crystallographic) rational Calogero model as its
hidden algebra \cite{GT}.  Note that the space ${\mathcal P}^{(1,2,3)}_{N}$ is invariant with respect to the quasi-projective transformation,
\begin{equation}
\label{QPT-tau}
     \ta_1 \rar \ta_1\ ,\  \ta_2 \rar \ta_2 + A \ta_1^2\ ,\ \ta_3 \rar \ta_3 + B \ta_1 \ta_2 + C \ta_1^3\ ,
\end{equation}
where $A, B, C$ are parameters. (AVT thanks M. Kontsevich for bringing attention to this property).

The algebra $h^{(3)}$ is infinite-dimensional but finitely-generated, for discussion see \cite{GT}. Their generating elements can be split into two classes. The first class of generators (lowering and Cartan
operators) act in $\mathcal{P}^{(1,2,3)}_N$ for any $N$ and
therefore they preserve the flag $\mathcal{P}^{(1,2,3)}$. The second
class operators (raising operators) act on the space
$\mathcal{P}^{(1,2,3)}_N$ with fixed $N$ only.

Let us introduce the following notation for the derivatives:
\[
\pa_i\equiv\frac{\pa}{\pa\tau_i}\ ,\quad
\pa_{ij}\equiv\frac{\pa^2}{\pa\tau_{i}\pa\tau_{j}}\
,\quad\pa_{ijk}\equiv\frac{\pa^3}{\pa\tau_{i}\pa\tau_{j}\pa\tau_{k}}\
.
\]
The first class of generating elements consists of the 22 generators where 13 of them are the first order operators
\begin{equation}
\begin{aligned}
\label{ops_1}
& T_0^{(1)}=\pa_1\,, && T_0^{(2)}=\pa_2\,, && T_0^{(3)}=\pa_3\,,\\
& T_1^{(1)}=\tau_1\pa_1\,, && T_2^{(2)}=\tau_2\pa_2\,, && T_3^{(3)}=\tau_3\pa_3\,,\\
& T_1^{(3)}=\tau_1\pa_3\,, && T_{11}^{(3)}=\tau_1^2\pa_3\,, && T_{111}^{(3)}=\tau_1^3\pa_3\,,\\
& T_1^{(2)}=\tau_1\pa_2\,, && T_{11}^{(2)}=\tau_1^2\pa_2\,, && T_2^{(3)}=\tau_2\pa_3\,,\\
& &&T_{12}^{(3)}=\tau_1\tau_2\pa_3\ ,&&
\end{aligned}
\end{equation}
the 6 are of the second order
\begin{equation}
\begin{aligned}
\label{ops_2}
& T_2^{(11)}=\tau_2\pa_{11}\,, && T_{22}^{(13)}=\tau_2^2\pa_{13}\,, && T_{222}^{(33)}=\tau_2^3\pa_{33}\,,\\
& T_3^{(12)}=\tau_3\pa_{12}\,, && T_3^{(22)}=\tau_3\pa_{22}\,, &&
T_{13}^{(22)}=\tau_1\tau_3\pa_{22}\ ,
\end{aligned}
\end{equation}
and 2 are of the third order
\begin{equation}
\begin{aligned}
\label{ops_3}
& T_3^{(111)}=\tau_3\pa_{111}\,, &&
T_{33}^{(222)}=\tau_3^2\pa_{222}\ .
\end{aligned}
\end{equation}

The generators of the second class consist of 8 operators where single one of them is of the first order
\begin{equation}
\label{R1}
T_1^+ = \ta_1 T_0\ ,
\end{equation}
4 are of the second order
\begin{equation}
\begin{aligned}
\label{R2}
& T_{2,-1}^+=\tau_2\pa_1T_0\,, &&
T_{3,-2}^+=\tau_3\pa_2T_0\,,  && T_{22,-3}^+ = \ta_2^2\pa_3T_0\,,
&& T_2^+ = \tau_2T_0(T_0+1)\ ,
\end{aligned}
\end{equation}
and 3 are of the third order
\begin{equation}
\begin{aligned}
\label{R3}
& T_{3,-11}^{+}=\tau_3\pa_{11}T_0\ , &&
T_{3,-1}^+=\tau_3\pa_1T_0(T_0+1)\ , &&
T_3^+=\tau_3T_0(T_0+1)(T_0+2)\ ,
\end{aligned}
\end{equation}
where we have introduced the diagonal operator (the Euler-Cartan generator)
\begin{equation}
\label{jo}
T_0=\tau_1\pa_1+2\tau_2\pa_2+3\tau_3\pa_3 - N\ .
\end{equation}
for a convenience. In fact, this operator is the identity operator,
it is of the zeroth order and, hence, it belongs to the first class.

It is not surprising that the algebraic operator $h^{(qes)}(\tau)$, (\ref{hQES-N-tau}), can be rewritten in terms of generators of the $h^{(3)}$-algebra,
\begin{equation}
\begin{aligned}
   h^{(qes)}(T) \ = & \  - \bigg[6\,T_1^{(1)}\,T_0^{(1)} + 2\,(7\,T_2^{(2)} - T_{11}^{(2)})\,T_1^{(2)} + T_3^{(3)}(6\,T_2^{(3)} - T_{11}^{(3)})
\\ &  \ +\ 12 T_0^{(1)}\,(2\,T_2^{(2)} +3\,T_3^{(3)}) + 2\,(4\,T_2^{(3)}\,T_2^{(2)}+9\,T_1^{(2)}\,T_3^{(3)}-\,T_{11}^{(3)}\,T_2^{(2)})
\\ &  + 2\,(9\,T_0^{(1)}+7\,T_1^{(2)}) + 2\,(7\,T_2^{(3)}-T_{11}^{(3)})
                          \bigg]
\label{hj}
\end{aligned}
\end{equation}
\[
 -\ 4\,(1\, + \,\gamma)\,(T_2^{(3)} + 2\,T_1^{(2)} + 3\,T_0^{(1)})\ +\ 12\,\om\,(T_0+N)\ +\ 12\,A\,T_1^+\ ,
\]
as well as the algebraic operator $h^{(es)}(\tau)$, (\ref{hES-N-tau}), which occurs at $A=0$, can be rewritten in terms of generators of the $h^{(3)}$-algebra,
\begin{equation}
\begin{aligned}
   h^{(es)}(T) \ = & \  -\bigg[6\,T_1^{(1)}\,T_0^{(1)}  +2\,(7\,T_2^{(2)} - T_{11}^{(2)})\,T_1^{(2)} + T_3^{(3)}(6\,T_2^{(3)} - T_{11}^{(3)})
\\ &  \ +\ 12 T_0^{(1)}\,(2\,T_2^{(2)}\, + \,3\,T_3^{(3)}) + 2\,(4\,T_2^{(3)}\,T_2^{(2)}+9\,T_1^{(2)}\,T_3^{(3)}-\,T_{11}^{(3)}\,T_2^{(2)})
\\ &  + 2\,(9\,T_0^{(1)}+7\,T_1^{(2)}) + 2\,(7\,T_2^{(3)}-T_{11}^{(3)})
                        \bigg]
\label{hj1}
\end{aligned}
\end{equation}
\[
   -\ 4\,(1\, + \,\gamma)\,(T_2^{(3)} + 2\,T_1^{(2)} + 3\,T_0^{(1)})\ +\ 12\,\om\,T_0\ ,
\]
where without a loss of generality we put $N=0$. The integral (\ref{integral-tau}) can be rewritten in terms of generators of the $h^{(3)}$-algebra as well,
\begin{equation}\label{integral-h3}
  -L_1^2 \ =\
  27 T_3^{(3)} T_3^{(3)} - 18 T_3^{(3)} T_{12}^{(3)} + 4 T_3^{(3)}T_{111}^{(3)} + 4 T_{222}^{(33)} - T_{12}^{(3)} T_{12}^{(3)} - 9 T_{12}^{(3)} + 2 T_{111}^{(3)}\ .
\end{equation}
It involves the generators of the first class only: (\ref{ops_1}), (\ref{ops_2}). Hence, it preserves the infinite flag of
polynomials ${\mathcal P}^{(1,2,3)}_{N}$, see (\ref{P3-tau}), $N=0,1,2, \ldots$.

It can be immediately verified that with respect to the action of the operator (\ref{hQES-N-tau}) the
finite-dimensional invariant subspace (\ref{P3-tau}) is reducible: it preserves
\begin{equation}
\label{P2-tau}
     {\mathcal P}^{(1,2)}_{N}\ \equiv \ \langle \ta_1^{p_1} \ta_2^{p_2} \vert \
     0 \le p_1+2p_2 \le N \rangle\ \subset {\mathcal P}^{(1,2,3)}_{N}\ .
\end{equation}
The operator which acts on ${\mathcal P}^{(1,2)}_{N}$ has the form,
\begin{equation}
\label{hQES-N-tau-2}
    h^{(qes)}(\ta_1,\ta_2) \ = \ -6\,\ta_1\pa_1^2 -2\ta_1(7\ta_2-\ta_1^2)\pa_2^2 -\,24\,\ta_2\pa_{1,2}^2\ -\
    6\,(5\, +\, 2\gamma)\pa_1 - 2(11\,+\,4\gamma)\ta_1\pa_2\
\end{equation}
\[
 +\ 12\om\,(\ta_1\pa_1+2\ta_2\pa_2)\ +\ 12\,A\ta_1(\ta_1\pa_1 + 2\ta_2\pa_2 - N) \ ,
\]
cf. (\ref{eq-psi}).
It has $\sim {N}^2$ polynomial eigenfunctions which depends on two variables $\ta_{1,2}$ only. The space ${\mathcal P}^{(1,2)}_{N}$ is finite-dimensional representation space of the non-semi-simple Lie algebra $g^{(2)} \subset gl(2,{\bf R}) \oplus R^3$ realized by the first order differential operators,
\cite{gko1} (see also \cite{Turbiner:1994}, \cite{ghko}, \cite{Turbiner:1998}),
\[ t_1\  =\  \pa_{\ta_1} \ , \]
\[ t_2 ({ N})\  =\ {\ta_1} \pa_{\ta_1}\ -\ \frac{{ N}}{3} \ ,
 \ t_3 ({ N})\  =\ 2 {\ta_2}\pa_{\ta_2}\ -\ \frac{{ N}}{3}\ ,\]
\[ t_4 ({ N})\  =\ {\ta_1}^2 \pa_{\ta_1} \  +\ 2 {\ta_1} {\ta_2} \pa_{\ta_2} \ - \ { N} {\ta_1} \ ,\]
\begin{equation}
\label{gr}
 r_{i}\  = \ {\ta_1}^{i}\pa_{\ta_2}\ ,\quad i=0, 1, 2\ .
\end{equation}
The operator (\ref{hQES-N-tau-2}) can be rewritten in terms of $gl(2,{\bf R}) \oplus R^3$ operators alone
\[
     h^{(qes)}(t,r)\ =\ -6\,r_1 t_1 - 14 (t_3 + \frac{{ N}}{3}) r_1 + 2 r_2 r_1 -  24 t_1 (t_3 + \frac{{ N}}{3})
\]
\[
     - 6(5 \,+\, 2\gamma)t_1 - 2(11\,+\,4\gamma) r_1 + 12\,\om (t_2 + t_3 + N)\ +\ 12 A t_4\ .
\]

The space (\ref{P3-tau}) is reducible further: the operator (\ref{hQES-N-tau}) (and also the operator (\ref{hQES-N-tau-2})) preserves
\begin{equation}
\label{P1-tau}
     {\mathcal P}^{(1)}_{N}\ \equiv \ \langle \ta_1^{p_1} \vert \
     0 \le p_1 \le N \rangle\ \subset {\mathcal P}^{(1,2)}_{N}\  \subset {\mathcal P}^{(1,2,3)}_{N}\ ,
\end{equation}
as well.
The operator, which acts on ${\mathcal P}^{(1)}_{N}$, has the form,
\begin{equation}
\label{hQES-N-tau-1}
    \frac{1}{6}h^{(qes)}(\ta_1) \ = \ -\,\ta_1\pa_1^2
  + \bigg(2\,A\ta_1^2 + 2\om\,\ta_1 - (5\, +\, 2\gamma)\bigg)\pa_1\ -
 2\,A N \ta_1 \ .
\end{equation}
It can be rewritten in terms of $sl(2, {\bf R})$ algebra generators,
\begin{equation}
\label{sl2}
    {\cal J}^+(N)\ =\ \ta_1^2\pa_{\ta_1} - N \ta_1\ ,\ {\cal J}^0(N)\ =\ 2\ta_1\pa_{\ta_1} - N\ ,\ {\cal J}^-(N)\ =\ \pa_{\ta_1}\ .
\end{equation}
It can be immediately recognized that the spectra of polynomial eigenfunctions of (\ref{hQES-N-tau-1}) corresponds to the spectra of the QES sextic polynomial potential with singular term $\sim 1/\ta_1$, see \cite{Turbiner:2016}, Case VII.

Eventually, it can be stated that among $\sim N^3$ polynomial eigenfunctions in $\ta$
variables  of the quasi-exactly-solvable operator
(\ref{hQES-N-tau}) there are $\sim N^2$ polynomial eigenfunctions of the quasi-exactly-solvable operator (\ref{hQES-N-tau-2}) and
$\sim N$ polynomial eigenfunctions of the quasi-exactly-solvable operator (\ref{hQES-N-tau-1}). A similar situation occurs for the
exactly-solvable operator (\ref{hES-N-tau}), see (\ref{hQES-N-tau}) at $A=0$, for which there exist infinitely-many polynomial
eigenfunctions in $\ta$
variables. Among these eigenfunctions there exists the infinite family of the polynomial eigenfunctions in $\ta_{1,2}$ variables, which
are eigenfunctions of the operator
\begin{equation}
\label{hES-N-tau-2}
    h^{(es)}(\ta_1,\ta_2) \ = \ -6\,\ta_1\pa_1^2 -2\ta_1(7\ta_2-\ta_1^2)\pa_2^2 -\,24\,\ta_2\pa_{1,2}^2\
    -\ 30\,\pa_1 - 22\,\ta_1\pa_2\ -\
\end{equation}
\[
 4\gamma(3\pa_1+2\ta_1\pa_2) +12\om\,(\ta_1\pa_1+2\ta_2\pa_2)\ .
\]
Besides that there exists the infinite family of the polynomial eigenfunctions in the $\ta_{1}$ variable, which are eigensolutions of the operator
\begin{equation}
\label{hES-N-tau-1}
    h^{(es)}(\ta_1) \ = \ -6\,\ta_1\pa_1^2\ +\ 6\,(2\,\om\,\ta_1\,-\,2\, \gamma\, -\, 5)\pa_1\ ;
\end{equation}
they are nothing but the Laguerre polynomials. The spectra of polynomial eigenfunctions is equidistant,
\[
    E_N\ =\ 12 \,\om\, N\ ,
\]
and it corresponds to the spectra of a harmonic oscillator (with a singular term $\sim 1/\ta_1$ in the potential).

Finally, we  emphasize that both the above-described QES problems in $\rho$ and $\ta$ variables exclude conceptually the limit $d=1$:
the determinant of the metric $g^{\mu\nu}(\rho)$ and $g^{\mu\nu}(\ta)$ is identically zero at $d=1$, since the area of the triangle of
interaction shrinks to zero, and the operator ${\De_{LB}}$ becomes singular.

\subsection{QES in geometrical variables for arbitrary $d$}

We consider the $n$-body Hamiltonian (\ref{Hrel-Mod})
\[
   {\tilde {\cal H}}_R\ = \ - {\De_R} + V_G\ ,
\]
written in geometrical variables $P, S, T$ and look for potentials $V_G$ for which there exists an (in)finite number of polynomial
eigenfunctions for any positive $d > 0$. This problem can be reduced to search for a square-integrable function $\Psi_0(P,\,S,\,T)$
such that the gauge-rotated operator $(\Psi_0)^{-1} \De_R \Psi_0$ remains algebraic (up to an additive function) and can be rewritten in
terms of the generators of the algebra $h^{(3)}$ (acting on functions of variables $P, S, T$).

\subsubsection{Exactly-solvable problem}

Let us take the operator (\ref{addition3-3tauS})
\[
\De_R \ =  \ 6\,P\,\pa^2_P + \frac{1}{2}\,P\,S\,\pa_{S}^2 +
   T\,(48\,S + P^2)\,\pa_{T}^2 + 36\,T\,\pa_{P,T} +
   24\,S\,\pa_{P,S} + 2\,S (16\,S + P^2)\,\pa_{S,T}
\]
\[
  +\ 6\,d\,\pa_P\ +\ \frac{1}{4}\,(d-1)\,P\,\pa_{S}\ +\ \frac{1}{2}\,[16\,(d+4)\,S + d\,P^2]\,\pa_{T}  \ ,
\]
and gauge-rotate it with a $T$-independent function
\begin{equation}
\label{psi_cal-d3-1}
       \Psi_0(P,\,S,\,T=0) \ = \
       {S}^{\tilde \gamma}\,e^{-\om\,P} \
       \equiv \Psi_0(P,S)\ ,\ \tilde \gamma\ \geq \ 0\ .
\end{equation}
As a result we get the additional terms to $\De_R$,
\[
   (\Psi_0)^{-1} \De_R \Psi_0\ =\ \De_R - 12\om (P\pa_P + 2 S \pa_S + 3 \,T\, \pa_{T})
   + 24 \tilde \gamma \pa_P + \tilde \gamma P \pa_S + 2 \tilde\gamma (16S + P^2) \pa_{T}
\]
\[
   + 6 \om^2 P + \frac{\tilde\gamma (2\tilde\gamma -3 + d)P}{4S} - 6\om(d + 4\tilde\gamma)
   \equiv h^{(exact)} + V^{(exact)} - E_0\ ,
\]
where evidently
\begin{equation}
\label{exact-III}
  h^{(exact)}\ =\ \ 6\,P\,\pa^2_P + \frac{1}{2}\,P\,S\,\pa_{S}^2 +
   T\,(48\,S + P^2)\,\pa_{T}^2 + 36\,\tau_3\,\pa_{P,T} +
   24\,S\,\pa_{P,S} + 2\,S (16\,S + P^2)\,\pa_{S,T}
\end{equation}
\[
   -12\om (P\pa_P + 2 S \pa_S + 3 T \pa_{T})
\]
\[
  +\ 6\,(d+4 \tilde\gamma)\,\pa_P\ +\ \frac{1}{4}\,(d-1+4 \tilde\gamma)\,P\,\pa_{S}\ +\ \frac{1}{2}\,[16\,(d+4+4 \tilde\gamma)\,S + (d+4 \tilde\gamma)\,P^2]\,\pa_{T}  \ ,
\]
is an exactly-solvable, algebraic operator, see below, and the potential
\begin{equation}
\label{V0-III}
    V^{(exact)}\ =\ 6 \om^2 P + \frac{\tilde\gamma (2\tilde\gamma -3 + d)}{4} \frac{P}{S}\ ,
\end{equation}
is the exactly-solvable many-body potential, which at $\tilde \gamma=0$ can be identified with a harmonic oscillator potential,
see e.g. \cite{Green}. The second terms play the role of a centrifugal potential due to rotation of the interaction plane (triangle)
around the center-of-mass. Here
\begin{equation}
\label{E0-III}
     E_0\ =\ 6\om(d + 4\tilde\gamma)\ ,
\end{equation}
is the ground state energy. The function $\Psi_0(P,S)$ is nothing but the ground state function for the potential (\ref{V0-III}); it is positive
in the configuration space $S_{\triangle}>0$.

It can be immediately checked that the exactly-solvable operator (\ref{exact-III}) has infinitely-many finite-dimensional invariant
subspaces in polynomials in variables $P, S, T$,
\begin{equation}
\label{P3-tauS}
     {\mathcal P}^{(1,2,3)}_{N}\ =\ \langle P^{p_1} S^{p_2} T^{p_3} \vert \
     0 \le p_1+2p_2+3p_3 \le N \rangle\ ,
\end{equation}
cf. (\ref{P3-tau}), with characteristic vector $(1,2,3)$, which form the infinite flag. The spectra of $-h^{(exact)}$ coincides with
the spectra of the Hamiltonian, it is
\begin{equation}
\label{E3-tauS}
               E_{p_1,p_2,p_3}\ =\ 12 \om (p_1 + 2p_2 + 3p_3 ) + E_0\ ,
\end{equation}
where $p_{1,2,3}=0,1,\ldots$ are quantum numbers, with multiplicity
\[
       M\ =\ p_1 + 2 p_2 + 3 p_3\ .
\]

The operator (\ref{exact-III}) acts on (\ref{P3-tauS}) reducibly. It maps
\[
   h^{(exact)}: {\mathcal P}^{(1,2,0)}_{N} \rar {\mathcal P}^{(1,2,0)}_{N} = \langle P^{p_1} S^{p_2} \vert \
     0 \le p_1+2p_2 \le N \rangle \equiv {\mathcal P}^{(1,2)}_{N} \subset {\mathcal P}^{(1,2,3)}_{N}\ .
\]
and
\[
   h^{(exact)}: {\mathcal P}^{(1,0,0)}_{N} \rar {\mathcal P}^{(1,0,0)}_{N} = \langle P^{p_1} \vert \
     0 \le p_1 \le N \rangle \equiv {\mathcal P}^{(1)}_{N} \subset {\mathcal P}^{(1,2,3)}_{N}\ .
\]
Therefore,  $h^{(exact)}$ preserves the subflag of spaces of polynomials made of ${\mathcal P}^{(1,2)}_{N},\ N=0,1,2,\ldots$ as
well as polynomials ${\mathcal P}^{(1)}_{N},\ N=0,1,2,\ldots$. It implies the existence of a (sub)-family of the eigenpolynomials of the
form $\mbox{Pol}_N (P, S)$ as well as another (sub)-family $\mbox{Pol}_N (P)$\,. The first sub-family leads to the eigenfunctions of the
reduced operator (\ref{exact-III}),
\begin{equation}
\label{exact-III-1}
  h^{(exact)}_r \ =\ \ 6\,P\,\pa^2_P + \frac{1}{2}\,P\,S\,\pa_{S}^2  + 24\,S\,\pa_{P,S} -12\om (P\pa_P + 2 S \pa_S)
\end{equation}
\[
  +\ 6\,(d+4 \tilde\gamma)\,\pa_P\ +\ \frac{1}{4}\,(d-1+4 \tilde\gamma)\,P\,\pa_{S}\ ,
\]
namely,
\[
   h^{(exact)}_r: {\mathcal P}^{(1,2)}_{N} \rar {\mathcal P}^{(1,2)}_{N}  \ .
\]
while the second sub-family leads to the eigenfunctions of another reduced operator
(\ref{exact-III}),
\begin{equation}
\label{exact-III-2}
  h^{(exact)}_{rr} \ =\ 6\,P\,\pa^2_P\ -\ 12\om \,P\pa_P \ +\ 6\,(d+4 \tilde\gamma)\,\pa_P\ ,
\end{equation}
namely,
\[
   h^{(exact)}_{rr}: {\mathcal P}^{(1)}_{N} \rar {\mathcal P}^{(1)}_{N}  \ .
\]
One can recognize that (\ref{exact-III-2}) is the Laguerre operator.

In general, the eigenfunctions of the algebraic sector of ${\tilde {\cal H}}_R\ = \ - {\De_R} + V^{(exact)}$, which correspond to
the eigenvalues (\ref{P3-tauS}) are factorized to the product of a polynomial and $\Psi_0 (P, S)$, thus, they are of the form
\[
   \mbox{Pol}_N (P, S, T)\ \Psi_0 (P, S)\ .
\]
However, among them there exist two particular forms of eigenfunctions,
\[
   \mbox{Pol}_N (P, S)\ \Psi_0 (P, S)\ ,
\]
and
\[
   \mbox{Pol}_N (P)\ \Psi_0 (P, S)\ .
\]

It is evident that they form the infinite family of eigenstates of the reduced $n$-body Hamiltonian ${\tilde {\cal H}}_R$, hence,
this problem is exactly solvable. We do not know whether their spectra is complete.
However, from the point of view of the original problem (\ref{Hrel}),
$${\cal H}_r =\ -\sum_{i=1}^3 \frac{1}{2 m_i} \De_i^{(d)}\ +\  V^{(exact)}\ ,$$
it is quasi-exactly-solvable, since it has infinitely-many angle-dependent eigenfunction which likely are of non-algebraic nature.

The limit $d=1$ corresponds to vanishing area of the interaction triangle, hence, $S=0$, and also $\tilde\gamma=0$. The ground state
function (\ref{psi_cal-d3-1}) becomes
\begin{equation}
\label{psi_cal-d3-d1}
       \Psi_0(P,\,S=0,\,T=0) \ =\ e^{-\om\,P} \ \equiv\ \Psi_0(P)\ ,\ \om > 0\ ,
\end{equation}
the operator $\De_R$ remains algebraic, see (\ref{addition3-3tauS2}), and also the operator (\ref{exact-III}),
\begin{equation}
\label{exact-III-d1}
  h^{(exact)}_{d=1}\ =\ \ 6\,P\,\pa^2_P + T\,P^2\,\pa_{T}^2 + 36\,T\,\pa_{P,T}
  - 12\om (P\pa_P + 3 T \pa_{T}) +\ 6\,\pa_P\ +\ \frac{P^2}{2}\,\pa_{T}  \ .
\end{equation}
It is easy to check that at $\om=0$ the operator $h^{(exact)}_{d=1}$ is the flat Laplace-Beltrami operator with metric (\ref{gmunu-d1}).
The potential contains no singular part,
\begin{equation}
\label{V0-III-d1}
    V^{(exact)}_{d=1}\ =\ 6 \om^2 P \ ,
\end{equation}
and coincides with regular part of the 3-body $G_2$ rational, Wolfes model.
The geometrical coordinates $P, T$ correspond to $\la_{1,2}$ (\ref{d1-la}), in those the 3-body $G_2$ rational, Wolfes model
becomes algebraic, and
\begin{equation}
\label{E0-III-d1}
     E_0{(d=1)}\ =\ 6\om\ .
\end{equation}

The spectra of the operator  $h^{(exact)}_{d=1}$ (\ref{exact-III-d1}) is equidistant
\begin{equation}
\label{P3-tauS-d0}
               E_{p_1,0,p_3}\ =\ 12 \om (p_1 + 3p_3 ) + 6 \om\ ,
\end{equation}
cf.(\ref{P3-tauS}), where $p_{1,3}=0,1,\ldots$ are quantum numbers, with multiplicity
\[
       M\ =\ p_1 + 3 p_3\ .
\]

It can be immediately checked that the exactly-solvable operator (\ref{exact-III-d1}) has infinitely-many finite-dimensional invariant subspaces of
polynomials in variables $P, T$,
\begin{equation}
\label{P3-tauS-d1}
     {\mathcal P}^{(1,0,3)}_{N}\ =\ \langle P^{p_1} T^{p_3} \vert \
     0 \le p_1+3p_3 \le N \rangle\ ,\ {\mathcal P}^{(1,0,3)}_{N} \subset {\mathcal P}^{(1,2,3)}_{N}
\end{equation}
cf. (\ref{P3-tauS}), with characteristic vector $(1,3)$, which form an infinite flag. It is easy to check that the operator
(\ref{exact-III-d1}) acts on (\ref{P3-tauS-d1}) reducibly. It maps
\[
   h^{(exact)}_{d=1}: {\mathcal P}^{(1,0,0)}_{N} \rar {\mathcal P}^{(1,0,0)}_{N} = \langle P^{p_1} \vert \
     0 \le p_1 \le N \rangle \equiv {\mathcal P}^{(1)}_{N} \subset {\mathcal P}^{(1,0,3)}_{N}\ .
\]

It leads to the eigenfunctions of a reduced operator
(\ref{exact-III-d1}),
\begin{equation}
\label{exact-III-d1-1}
  h^{(exact)}_{d=1,r} \ =\ 6\,P\,\pa^2_P\ -\ 12\om \,P\pa_P \ +\ 6\,\pa_P\ ,
\end{equation}
cf. (\ref{exact-III-2}). It is again the Laguerre operator.

In general, the eigenfunctions of the algebraic sector of ${\tilde {\cal H}}_R\ = \ - {\De_R} + V^{(exact)}_{d=1}$,
which correspond to the eigenvalues (\ref{P3-tauS}-d1) are factorized as the product of a polynomial and $\Psi_0 (P)$, thus, they are of the form
\[
   \mbox{Pol}_N (P, T)\ \Psi_0 (P)\ .
\]
However, among them there exists a particular form of eigenfunctions,
\[
   \mbox{Pol}_N (P)\ \Psi_0 (P)\ .
\]

\subsubsection{Quasi-Exactly-solvable problem}

Let us take the function
\begin{equation}
\label{psi_cal-d3-2}
       \Psi_0(P,\,S,\,T=0) \ = \
       {S}^{\tilde \gamma}\,e^{-\om\,P - \frac{A}{2} P^2} \
       \equiv \Psi_0(P,S)\ ,\ \tilde \gamma\ \geq \ 0\ ,\ A \geq 0\ ,
\end{equation}
cf.(\ref{psi_cal-d3-1}), and make the gauge rotation of $\De_R$ (\ref{addition3-3tauS})
with $\Psi_0$. As the result we get $\De_R$ and the additional first order terms, overall it can be split into three terms
\[
   (\Psi_0)^{-1} \De_R \Psi_0\ =\ \De_R - 12\om (P\pa_P + 2 S \pa_S + 3 T \pa_{T})
   + 24 \tilde \gamma \pa_P + \tilde \gamma P \pa_S + 2 \tilde\gamma (16S + P^2) \pa_{T}
\]
\[
   + 6 \om^2 P + \frac{\tilde\gamma (2\tilde\gamma -3 + d)P}{4S} - 6\om(d + 4\tilde\gamma)
   \equiv h^{(qes)} + V^{(qes)} - E_0\ ,
\]
where
\begin{equation}
\label{qes-III}
  h^{(qes)}\ =\ \ 6\,P\,\pa^2_P + \frac{1}{2}\,P\,S\,\pa_{S}^2 +
   T\,(48\,S + P^2)\,\pa_{\ta_3}^2 + 36\,T\,\pa_{P,T} +
   24\,S\,\pa_{P,S} + 2\,S (16\,S + P^2)\,\pa_{S,T}
\end{equation}
\[
   -12\om (P\pa_P + 2 S \pa_S + 3 T \pa_{T})
\]
\[
  +\ 6\,(d+4 \tilde\gamma)\,\pa_P\ +\ \frac{1}{4}\,(d-1+4 \tilde\gamma)\,P\,\pa_{S}\ +\ \frac{1}{2}\,[16\,(d+4+4 \tilde\gamma)\,S + (d+4 \tilde\gamma)\,P^2]\,\pa_{T}
\]
\[
   - 12 A P (P\pa_P + 2 S \pa_S + 3 T \pa_{T} - N)  \ ,
\]
is the algebraic, quasi-exactly-solvable, if $N$ is integer, operator, see below, and the potential
\begin{equation}
\label{V0-qes-III}
    V^{(qes)}\ =\ 6 [\om^2 - A (4 \tilde\gamma + 2 N + d + 1)]\,P + 12 \om A P^2  + 6 A^2 P^3 + \frac{\tilde\gamma (2\tilde\gamma -3 + d)}{4} \frac{P}{S}\ ,
\end{equation}
is the quasi-exactly-solvable many-body sextic potential, which at $A=\tilde \gamma=0$ can be identified with harmonic oscillator (non-singular) potential (\ref{V0-III}), see e.g. \cite{Green}, and $E_0$ (\ref{E0-III}).
The last term in $V^{(qes)}$ plays the role of a centrifugal potential due to rotation of the interaction plane (triangle) around the center-of-mass (baricenter). Note that the term $(2 A N P)$ is added to (\ref{qes-III}) and subtracted in (\ref{V0-qes-III}).

For general value of the parameter $N$ the operator (\ref{qes-III}) is $h^{(3)}$ Lie-algebraic: it can be rewritten in terms of the generators of the algebra $h^{(3)}$ (\ref{ops_1})-(\ref{ops_2}), (\ref{R1}).
However, it can be immediately checked that for integer $N$ the operator (\ref{qes-III}) has single finite-dimensional invariant subspace in polynomials in variables $P, S, T$,
\[
     {\mathcal P}^{(1,2,3)}_{N}\ =\ \langle P^{p_1} S^{p_2} T^{p_3} \vert \
     0 \le p_1+2p_2+3p_3 \le N \rangle\ ,
\]
cf. (\ref{P3-tauS}), with characteristic vector $(1,2,3)$. Thus, it is quasi-exactly-solvable operator where $\sim N^3/6$ eigenstates can be found algebraically. In particular, it can be constructed the algebraic secular equation of the degree $\sim N^3/6$ with real roots alone whose roots are the eigenvalues. Simple analysis of the operator (\ref{qes-III}) shows that among the $\sim N^3/6$ polynomial eigenfunctions in three variables $P,S, T$ there exist the $\sim N^2/2$ polynomial eigenfunctions in two variables $P,S$ and the $(N+1)$ polynomial eigenfunctions in variable $P$. These latter eigenfunctions are the eigenfunctions of the operators
\begin{equation}
\label{III-qes-1}
  h^{(qes)}_r \ =\ \ 6\,P\,\pa^2_P + \frac{1}{2}\,P\,S\,\pa_{S}^2  + 24\,S\,\pa_{P,S} -12\om (P\pa_P + 2 S \pa_S)
\end{equation}
\[
  +\ 6\,(d+4 \tilde\gamma)\,\pa_P\ +\ \frac{1}{4}\,(d-1+4 \tilde\gamma)\,P\,\pa_{S} - 12 A P (P\pa_P + 2 S \pa_S - N)  \ ,
\]
cf. (\ref{exact-III-1}) and
\begin{equation}
\label{III-qes-2}
  h^{(qes)}_{rr} \ =\ 6\,P\,\pa^2_P\ -\ 12\om \,P\pa_P \ +\ 6\,(d+4 \tilde\gamma)\,\pa_P- 12 A P (P\pa_P - N)  \ ,
\end{equation}
cf. (\ref{exact-III-2}), respectively. It is easy to check that the quasi-exactly-solvable operator $h^{(qes)}_r$ (\ref{III-qes-1}) can be rewritten in terms of the generators of the algebra $g^{(2)} \supset gl(2, {\bf R}) \ltimes {\cal R}^{(2)}$, see e.g. \cite{Turbiner:1998},\cite{TTW:2009} while the (\ref{III-qes-2}) can be rewritten in terms of the generators of the algebra $sl(2)$ see \cite{Turbiner:1988}. The quasi-exactly-solvable operator $h^{(qes)}_{rr}$ corresponds to the Case VII in classification of one-dimensional QES operators \cite{Turbiner:2016} and describes the algebraic sector of the one-dimensional QES singular sextic polynomial potential. Note that the ground state function of the QES $n$-body Hamiltonian (\ref{Hrel-Mod}) with potential (\ref{V0-qes-III})
\[
   {\tilde {\cal H}}_R\ = \ - {\De_R}\ +\ V^{(qes)}\ ,
\]
is of the form
\[
  \Psi_{ground\, state}\ =\ P_N(P) \Psi_0(P,S)\ =\ P_N(P)\,{S}^{\tilde \gamma}\,e^{-\om\,P - \frac{A}{2} P^2}
  \ ,
\]
see (\ref{psi_cal-d3-2}), where $P_N(P)$ is the positive eigenfunction of the operator $h^{(qes)}_{rr}$ at $P > 0$.

In the limit $d=1$ the area of the interaction triangle vanishes $S=0$ as well as $\tilde\gamma=0$ to ensure that the ground state function (\ref{psi_cal-d3-2}) remains finite
\begin{equation}
\label{psi_qes-d1}
       \Psi_0(P,\,S=0,\,T=0) \ =\ e^{-\om\,P\ -\ \frac{A}{2}\, P^2} \
       \equiv\ \Psi_0(P)\ ,\ \om \geq 0\ ,\ A \geq 0\ .
\end{equation}
the operator $h^{(qes)}$ remains algebraic,
\begin{equation}
\label{qes-III-d1}
  h^{(qes)}_{d=1}\ =\ \ 6\,P\,\pa^2_P +
   T\,P^2\,\pa_{T}^2 + 36\,T\,\pa_{P,T} +\ 6\,\pa_P\ +\ \frac{P^2}{2}\,\pa_{T}
\end{equation}
\[
   -12\om (P\pa_P + 3 T \pa_{T}) - 12 A P (P\pa_P + 3 T \pa_{T} - N)  \ .
\]
It is easy to check that at $\om=0, A=0$ the operator $h^{(qes)}_{d=1}$ is the flat Laplace-Beltrami operator with metric (\ref{gmunu-d1}). The operator $h^{(qes)}_{d=1}$ (\ref{qes-III-d1}) can be rewritten in terms of the generators of the algebra $g^{(3)} \supset gl(2, {\bf R}) \ltimes {\cal R}^{(3)}$, see e.g. \cite{TTW:2009}. If $N$ is integer, the operator $h^{(qes)}_{d=1}$ has finite-dimensional invariant subspace
\[
     {\mathcal P}^{(1,3)}_{N}\ =\ \langle P^{p_1} T^{p_3} \vert \
     0 \le p_1+3p_3 \le N \rangle\ ,
\]
its $\sim N^2/2$ eigenfunctions are $N$th degree polynomials in variables $P, T$. Interestingly,
among these eigenfunctions there are $(N+1)$ eigenfunctions in the form of polynomials of degree $N$ in variable $P$. They are the eigenfunctions of the operator
\begin{equation}
\label{III-qes-2-d1}
  h^{(qes)}_{d=1,r} \ =\ 6\,P\,\pa^2_P\ -\ 12\om \,P\pa_P \ +\ 6\,\pa_P - 12 A P (P\pa_P - N)  \ ,
\end{equation}
cf.(\ref{qes-III-d1}). The quasi-exactly-solvable operator $h^{(qes)}_{r,d=1}$ corresponds to the Case VI in classification of one-dimensional QES operators \cite{Turbiner:2016} and describes the algebraic sector of the one-dimensional QES (non-singular) sextic polynomial potential. It can be rewritten in terms of the generators of the algebra $sl(2, {\bf R} )$.

The potential of the QES $n$-body Hamiltonian (\ref{Hrel-Mod}) at $d=1$ contains no singular part,
\begin{equation}
\label{V0-qes-III-d1}
    V^{(qes)}\ =\ 6 [\om^2 - 2 A (N + 1)]\,P + 12 \om A P^2  + 6 A^2 P^3 \ .
\end{equation}
Its ground state has the form
\[
  \Psi_{ground \,state,\, d=1}\ =\ P_N(P)\,e^{-\om\,P - \frac{A}{2} P^2}  \ ,
\]
where $P_N(P)$ is the lowest eigenfunction of the operator $(-h^{(qes)}_{d=1,r})$ with the property
$P_{N}(P) > 0$ at $P>0$.

\subsection{Primitive QES problems}

{\bf (a)} Let us take the $S_3$-permutationally symmetric function

\begin{equation}
\label{psi_cal}
       \Psi_a (r_{12},\,r_{13},\,r_{23})  \ = \  {(r_{12}\,r_{13}\,r_{23})}^{\gamma}\,e^{-\frac{\om}{2}(r_{12}^2\, + \, r_{13}^2\, +  \, r_{23}^2)} \ = \
       {\ta_3}^{\frac{\gamma}{2}}\,e^{-\frac{\om}{2}\,\ta_1} \ ,
\end{equation}
where $\gamma,\,\om > 0$ are constants and $\ta$'s are given by (\ref{taus}). If $d=1$, then, for the ordering $r_1 \leq r_2 \leq r_3$,
\begin{equation}
\label{con}
r_{23}=|r_{12}-r_{13}| \ ,
\end{equation}
and (\ref{psi_cal}) becomes 3-body Calogero ground state function (the Wigner-Dyson distribution). Here, (\ref{psi_cal}) is a natural generalization to arbitrary $d$.

Now we look for the potential for which the expression (\ref{psi_cal}) is the
ground state function for the Hamiltonian ${\cal H}_{r}$, see (\ref{Hrel}), (\ref{Hrel-Mod}).
This potential can be found immediately by calculating the ratio
\[
\frac{\De_{R}(r) \Psi_a}{ \Psi_a}\ =\ V_a - E_a \ ,
\]
where $\De_{R}(r)$ is given by (\ref{addition3-3r}).
The result is
\[
V_a^{(d)} \ = \   2\,\gamma\,(d+2\,\gamma-2)\bigg[\frac{1}{r_{12}^2} +  \frac{1}{r_{13}^2} +  \frac{1}{r_{23}^2} \bigg] - \gamma^2\,\bigg[ \frac{r_{12}^2}{r_{13}^2\,r_{23}^2} +  \frac{r_{13}^2}{r_{12}^2\,r_{23}^2}  +  \frac{r_{23}^2}{r_{12}^2\,r_{13}^2}  \bigg]
\]
\begin{equation}
+ 3\,\om^2\,(r_{12}^2+r_{13}^2+r_{23}^2)\ ,
\label{VQES2-0}
\end{equation}
with the energy of the ground state
\begin{equation}
  E_a\ =\ 6\,\om\,(d+3\,\gamma) \ .
\label{EQES2-0}
\end{equation}
It can be checked that for $d=1$, imposing (\ref{con}), the potential (\ref{VQES2-0}) becomes the familiar 3-body Calogero potential \cite{Calogero:1969},
\[
V_a^{(d=1)} \ = \   2\,\gamma\,(\gamma-1)\bigg[ \frac{1}{r_{12}^2} +  \frac{1}{r_{13}^2} +  \frac{1}{{(r_{13}-r_{12})}^2} \bigg] + 6\,\om^2\,(r_{12}^2+r_{13}^2-r_{12}\,r_{13})\ .
\]

Let us define the Hamiltonian
\[
 {\cal H}^{(a)}_{r} \ =\ -{ \De_R} + V_a^{(d)} \ ,
\]
and make a gauge rotation $\psi_a^{-1}\,{\cal H}^{(a)}_{r}\,\psi_a\,=\,-\De^{\prime}_R $,

\begin{equation}
\label{DRr}
    \De^{\prime}_R(r)\ =\ \ 2\,(\pa^{2}_{r_{12}} +\pa^{2}_{r_{23}}+\pa^{2}_{r_{13}})
\end{equation}
\[
\  +\ \frac{r_{12}^2-r_{13}^2+r_{23}^2}{r_{12} r_{23}}\,\pa_{r_{12}}\pa_{r_{23}}\  +\ \frac{r_{12}^2+r_{13}^2-r_{23}^2}{r_{12} r_{13}}\,\pa_{r_{12}}\pa_{r_{13}}\ + \ \frac{r_{13}^2+r_{23}^2-r_{12}^2}{r_{13} r_{23}}\,\pa_{r_{23}}\pa_{r_{13}} \
\]
\[
+ \ \frac{2(d-1)(r_{13}^2\,r_{23}^2)+\gamma\,(6r_{13}^2\,r_{23}^2+r_{12}^2(r_{13}^2+r_{23}^2)-r_{13}^4-r_{23}^4)\,
-6\,\om\,r_{12}^2r_{13}^2r_{23}^2}{r_{12}\,r_{13}^2\,r_{23}^2}\,\pa_{r_{12}}
\]
\[
+ \ \frac{2(d-1)(r_{13}^2\,r_{12}^2)+\gamma\,(6r_{13}^2\,r_{12}^2+r_{23}^2(r_{13}^2+r_{12}^2)-r_{13}^4-r_{12}^4)\,
-6\,\om\,r_{12}^2r_{13}^2r_{23}^2}{r_{23}\,r_{13}^2\,r_{12}^2}\,\pa_{r_{23}}
\]
\[
+ \ \frac{2(d-1)(r_{12}^2\,r_{23}^2)+\gamma\,(6r_{12}^2\,r_{23}^2+r_{13}^2(r_{12}^2+r_{23}^2)-r_{12}^4-r_{23}^4)\,
-6\,\om\,r_{12}^2r_{13}^2r_{23}^2}{r_{13}\,r_{12}^2\,r_{23}^2}\,\pa_{r_{13}}\ .
\]

By construction the operator $\De^{\prime}_R(r)$ (\ref{DRr}) has a single one-dimensional invariant subspace $<1>$ in space of polynomials mapping it to itself. Hence, the Hamiltonian ${\cal H}^{(a)}_{r}$ is a
primitive QES problem where only the ground state is known.

{\bf (b)} Let us take another $S_3$-permutationally symmetric function

\begin{equation}
\label{psi_cal-b}
       \Psi_b (r_{12},\,r_{13},\,r_{23})  \ = \  {(|r_{12}-r_{13}|\,|r_{13}-r_{23}|\,|r_{12}-r_{23}|)}^{\gamma}\,e^{-\frac{\om}{2}(r_{12}^2\, + \, r_{13}^2\, +  \, r_{23}^2)} \ ,
\end{equation}
cf. (\ref{psi_cal}), where $\gamma,\,\om > 0$ are constants. If $d=1$, then, for the ordering $r_1 \leq r_2 \leq r_3$,
\begin{equation}
r_{23}=|r_{12}-r_{13}| \ ,
\end{equation}
and (\ref{psi_cal-b}) becomes 3-body Calogero ground state function as (\ref{psi_cal}) does. Also (\ref{psi_cal-b}) is a natural generalization to arbitrary $d$.

The potential for which the expression (\ref{psi_cal-b}) is the ground state function for
the Hamiltonian ${\cal H}_{r}$, see (\ref{Hrel}) and (\ref{Hrel-Mod}), is given by
\[
V_b^{(d)} \ = \ \frac{\gamma ^2 \left({\si}_1^7-9 {\si}_1^5 {\si}_2 + 33 {\si}_1^4 {\si}_3 +20 {\si}_1^3 {\si}_2^2 -153 {\si}_1^2 {\si}_2 {\si}_3 +162 {\si}_1 {\si}_3^2 +54 {\si}_2^2 {\si}_3\right)}{{\si}_3 \left( 18 {\si}_1 {\si}_2 {\si}_3 +{\si}_1^2 {\si}_2^2 -4  {\si}_1^3 {\si}_3 -4 {\si}_2^3-27 {\si}_3^2 \right)}\ +\ 3\om^2\, \left({\si}_1^2-2 {\si}_2\right) \ +
\]
\begin{equation}
\frac{\gamma  [54 (d-2) {\si}_1 {\si}_3^2+{\si}_3 \left((8d-25){\si}_1^4-9 (4 d-13) {\si}_2 {\si}_1^2-54 {\si}_2^2\right)-{\si}_1 ({\si}_1^2-4 {\si}_2) (2 (d+1) {\si}_2^2+{\si}_1^4-5 {\si}_2 {\si}_1^2)]}{{\si}_3 ( 18 {\si}_1 {\si}_2 {\si}_3 +{\si}_1^2 {\si}_2^2 -4  {\si}_1^3 {\si}_3 -4 {\si}_2^3-27 {\si}_3^2)} \ ,
\label{VQES3-0}
\end{equation}
where
\begin{equation*}
\begin{aligned}
&{\si}_1 \ = \ r_{12} + r_{13} +  r_{23} \ ,
\\ &  {\si}_2 \ = \ r_{12}\,r_{13} + r_{12}\,r_{23} +r_{13}\,r_{23} \ ,
\\ & {\si}_3 \ = \ r_{12}\,r_{13}\,r_{23} \ ,
\end{aligned}
\end{equation*}
are $S^3$ permutationally-symmetric, relative $r$-coordinate polynomials (elementary symmetric
polynomials in $r_{ij}$), see (\ref{d-si}), with the energy of the ground state
\begin{equation}
  E_b\ =\ 6\,\om\,(d+3\,\gamma) \ .
\label{EQES3-0}
\end{equation}
It can be checked that for $d=1$ imposing (\ref{con}) the potential (\ref{VQES3-0}) becomes the familiar 3-body Calogero potential
\cite{Calogero:1969},
\[
V_b^{(d=1)} \ = \   2\,\gamma\,(\gamma-1)\bigg[ \frac{1}{r_{12}^2} +  \frac{1}{r_{13}^2} +  \frac{1}{{(r_{13}-r_{12})}^2} \bigg] +
6\,\om^2\,(r_{12}^2+r_{13}^2-r_{12}\,r_{13})\ .
\]

Let us define the Hamiltonian
\[
 {\cal H}^{(b)}_{r} \ =\ -{ \De_R} + V_b^{(d)} \ ,
\]
and make a gauge rotation $\psi_b^{-1}\,{\cal H}^{(b)}_{r}\,\psi_b\,=\,-\De^{\prime}_R$,

\begin{equation}
\label{DRb}
   \De^{\prime}_R(r)\ =\ \ 2\,(\pa^{2}_{r_{12}} +\pa^{2}_{r_{23}}+\pa^{2}_{r_{13}})
\end{equation}
\[
\  +\ \frac{r_{12}^2-r_{13}^2+r_{23}^2}{r_{12} r_{23}}\,\pa_{r_{12}}\pa_{r_{23}}\  +\ \frac{r_{12}^2+r_{13}^2-r_{23}^2}{r_{12} r_{13}}\,\pa_{r_{12}}\pa_{r_{13}}\ + \ \frac{r_{13}^2+r_{23}^2-r_{12}^2}{r_{13} r_{23}}\,\pa_{r_{23}}\pa_{r_{13}} \
\]
\[
-\ \bigg[   \gamma\,\frac{ 8r_{12}r_{13}r_{23}(r_{13}+r_{23})  + r_{12}^4 +{(r_{13}+r_{23})}^4 -5r_{13}r_{23}(  r_{12}^2 +{(r_{13}+r_{23})}^2  ) -2r_{12}^2{(r_{13}+r_{23})}^2  }{r_{12}\,r_{13}\,r_{23}\,(r_{12}-r_{13})(r_{12}-r_{23})}
\]
\[
  - \frac{2\,(d-1)}{r_{12}}+6\,\om\,r_{12} \bigg]   \,\pa_{r_{12}}
\]

\[
-\ \bigg[   \gamma\,\frac{ 8r_{12}r_{13}r_{23}(r_{12}+r_{23})  + r_{13}^4 +{(r_{12}+r_{23})}^4 -5r_{12}r_{23}(  r_{13}^2 +{(r_{12}+r_{23})}^2  ) -2r_{13}^2{(r_{12}+r_{23})}^2  }{r_{12}\,r_{13}\,r_{23}\,(r_{13}-r_{12})(r_{13}-r_{23})}
\]
\[
  - \frac{2\,(d-1)}{r_{13}}+6\,\om\,r_{13} \bigg]   \,\pa_{r_{13}}
\]

\[
-\ \bigg[   \gamma\,\frac{ 8r_{12}r_{13}r_{23}(r_{13}+r_{12})  + r_{23}^4 +{(r_{13}+r_{12})}^4 -5r_{13}r_{12}(  r_{23}^2 +{(r_{13}+r_{12})}^2  ) -2r_{23}^2{(r_{13}+r_{12})}^2  }{r_{12}\,r_{13}\,r_{23}\,(r_{23}-r_{13})(r_{23}-r_{12})}
\]
\[
  - \frac{2\,(d-1)}{r_{23}}+6\,\om\,r_{23} \bigg]   \,\pa_{r_{23}}  \ .
\]

The operator (\ref{DRb}) has no invariant subspaces except for $<1>$. Hence, the Hamiltonian ${\cal H}^{(b)}_{r}$ is also a primitive
QES problem where only the ground state is known.

\section*{Conclusions}

In this paper for the 3-body problem with equal masses in d-dimensional space it is found the Schr\"odinger type equation in the space ${\bf \tilde R}$ of relative distances $\{ r_{ij} \}$,
\begin{equation}
\label{H3}
     {H}_{LB}\, \Psi(r_{12},\,r_{13},\,r_{23}) \ =\ E\, \Psi(r_{12},\,r_{13},\,r_{23})\ ,\qquad
    {H}_{LB} \ =\ -\De_{LB}(r_{ij}) +  \tilde V(r_{ij};d) +  V(r_{ij}) \ ,
\end{equation}
where the Laplace-Beltrami operator $\De_{LB}$, see e.g. (\ref{LB3}), is $d$-independent and makes sense as the  kinetic energy of a three-dimensional particle in curved space with metric (\ref{gmn33-rho})
for $d>1$, and for $d=1$ it degenerates to the kinetic energy of a two-dimensional particle in flat space, and $\tilde V(r_{ij};d)$ is effective potential. The operator (\ref{H3}) describes angle-independent solutions of the original 3-body problem (\ref{Hgen}),
including the ground state. Hence, finding the ground state (and some other states) involves the solution of the differential equation in three variables, contrary to the original $(2d)$-dimensional Schr\"odinger equation of the relative motion. Since the Hamiltonian ${H}_{LB}$ is Hermitian, the variational method
can be employed with only three-dimensional integrals involved. Generalization to the case of three bodies of arbitrary masses is straightforward and is done in the Appendix. The classical analogue of the Hamiltonian in (\ref{H3}) was presented as well in (\ref{Hrel-Cl-final}). Note that in this case the operator $\De_R$ is algebraic in the $\rho$-representation but not in $\tau-$representation or geometric variables representation.

The gauge-rotated Laplace-Beltrami operator, with $d$-independent determinant of the metric $D$ raised to a certain degree as the gauge factor, appears as an algebraic operator both in the variables which are squares of relative distances and which are the elementary symmetric polynomials in squares of relative distances as arguments. The former algebraic operator has the hidden algebra $sl(4, \bf R)$, while latter one has the hidden algebra $h^{(3)}$, thus, becoming Lie-algebraic operators. Both operators can be extended to (quasi)-exactly-solvable operators with potentials in a form of rational functions in either variables. Interestingly, both (quasi)-exactly-solvable operators (with hidden algebra $sl(4, \bf R)$ and $h^{(3)}$, respectively) lead to the {\it same} (quasi)-exactly-solvable potentials in the space of relative distances. Naturally, these (quasi)-exactly-solvable potentials in the space of relative distances are quasi-exactly-solvable potentials in the space of relative motion. We show there exists a special (quasi)-exactly-solvable problem in geometric variables $P,S,T$ which admits limit $d=1$ preserving polynomiality of the eigenfunctions. The ground state function always depends on the single variable $P$.
The exactly-solvable problem looks as the singular harmonic oscillator in the space of relative distances,
while the quasi-exactly-solvable problem appears as the singular sextic anharmonic oscillator. Both problems will be discussed in details elsewhere.

\section*{Acknowledgments}

A.V.T. is thankful to University of Minnesota, USA for kind hospitality extended to him where this work was initiated and IHES, France where it was mostly completed. He is deeply grateful to I.E.~Dzyaloshinsky, T.~Damour and M.~Kontsevich for useful discussions and important remarks in the early stage of the work.
A.V.T. is supported in part by the PAPIIT grant {\bf IN108815}.
W.M. was partially supported by a grant from the Simons Foundation (\# 412351 to Willard Miller, Jr.).
M.A.E.R. is grateful to ICN UNAM, Mexico for the kind hospitality during his visit, where a part of the research was done as well as CRM, Montreal, where it was completed, he was supported in part by DGAPA grant {\bf IN108815} (Mexico) and, in general, by  CONACyT grant {\bf 250881}~(Mexico) for postdoctoral research.

\section*{Appendix: non-equal masses}

Consider the general case of the particles located at points ${\bf r}_1,{\bf r}_2,{\bf r}_3$
of masses $m_1,m_2,m_3$, respectively.
Then the operator (\ref{addition3-3rho}) becomes (in terms of the relative coordinates $\rho_{ij}=r_{ij}^2$):
\[
  {\De_R}'(\rho_{ij})\ =\
  \frac{2}{\mu_{13}} \rho_{13}\, \pa_{\rho_{13}}^2 +
  \frac{2}{\mu_{23}} \rho_{23}\, \pa_{\rho_{23}}^2 +
  \frac{2}{\mu_{12}} \rho_{12}\,\pa_{\rho_{12}}^2 +
\]
\[
  \frac{2(\rho_{13} + \rho_{12} - \rho_{23})}{m_1}\pa_{\rho_{13}\rho_{12}} +
  \frac{2(\rho_{13} + \rho_{23} - \rho_{12})}{m_3}\pa_{\rho_{13}\rho_{23}} +
  \frac{2(\rho_{23} + \rho_{12} - \rho_{13})}{m_2}\pa_{\rho_{23}\rho_{12}} +
\]
\begin{equation}
\label{addition3-3r-M}
  \frac{d}{\mu_{13}} \pa_{\rho_{13}} +
  \frac{d}{\mu_{23}} \pa_{\rho_{23}} +
  \frac{d}{\mu_{12}} \pa_{\rho_{12}}\ ,
\end{equation}
where
\[
   \frac{1}{\mu_{ij}}\ =\ \frac{m_i+m_j}{m_i m_j}\ ,
\]
is reduced mass for particles $i$ and $j$; it is in agreement with  (\ref{addition3-3rho})
for $m_1=m_2=m_3=1$. At $d=3$ it coincides with (68) at \cite{{TMA:2016}}. This operator has the same algebraic structure as ${\De_R}(\rho_{ij})$
but lives on a different manifold in general. It can be rewritten in terms of the
generators of the maximal affine subalgebra $b_4$ of the algebra $sl(4,{\bf R})$,
see (\ref{sl4R}), c.f. (\ref{HRex}). The contravariant metric tensor is does not depends on $d$ and
its determinant is
\[
  D_m\ =\ \det g^{\mu \nu}\ =\ 2\,\frac{m_1+m_2+m_3}{m_1^2m_2^2m_3^2} \times
\]
\begin{equation}
\label{gmn33-rho-det-M}
 \left(m_1m_2\rho_{12}+m_1m_3\rho_{13}+m_2m_3\rho_{23}\right)
                     \left(2\rho_{12}\rho_{13} + 2 \rho_{12}\rho_{23} + 2 \rho_{13}\rho_{23}-\rho_{12}^2- \rho_{13}^2 - \rho_{23}^2\right) \ ,
\end{equation}
and is positive definite. It is worth noting a remarkable factorization property of the determinant
\[
D_m\ =\ 2\frac{m_1+m_2+m_3}{m_1^2m_2^2m_3^2} \,(m_1 m_2 r_{12}^2+m_1 m_3 r_{13}^2+m_2 m_3 r_{23}^2)\ \times
\]
\[
(r_{12}+r_{13}-r_{23})(r_{12}+r_{23}-r_{13})(r_{13}+r_{23}-r_{12})(r_{12}+r_{13}+r_{23})\ =
\]
\[
   =\ 32\, \frac{m_1+m_2+m_3}{m_1^2m_2^2m_3^2}\, P_m \ S^2_{\triangle}\ ,
\]
where $P_m=m_1 m_2 r_{12}^2+m_1 m_3 r_{13}^2+m_2 m_3 r_{23}^2$ - the weighted sum of squared of sides of the interaction triangle and ${S}_{\triangle}$ is their area. Hence, $D_m$ is still proportional to  ${S}_{\triangle}^2$, c.f. (\ref{gmn33-rho-det-gamma}).

Making the gauge transformation of (\ref{addition3-3r-M}) with determinant (\ref{gmn33-rho-det-M}) as the factor,
\[
         \Gamma \ = \ D_m^{-1/4} (4 \ta_2-\ta_1^2)^{\frac{3-d}{4}}\ =\ D_m^{-1/4}\,(16 S^2_{\triangle})^{\frac{3-d}{4}} \sim (P_m)^{-1/4} (S^2_{\triangle})^{\frac{2-d}{4}} \ ,
\]
we find that
\begin{equation}
         \Gamma^{-1}\, {\De_R}'(\rho_{ij})\,\Gamma \ =
        \  \De'_{LB}(\rho_{ij}) - \tilde V_m(\rho_{ij}) \ ,
\label{HLB3M}
\end{equation}
is the Laplace-Beltrami operator with the effective potential
\[
{\tilde V_m} \ =\ \frac{3}{8}\ \frac{(m_1+m_2+m_3)}{\left(m_1 m_2 \rho_{12}+m_1 m_3 \rho_{13}+m_2 m_3 \rho_{23}\right)}\ -
\]
\[
 \frac{(d-2)(d-4)}{2}\ \frac{\left(m_1 m_2 \rho_{12}+m_1 m_3 \rho_{13}+m_2 m_3 \rho_{23}\right)}
  { m_1 m_2 m_3\left(\rho_{12}^2+\rho_{13}^2+\rho_{23}^2 -2 \rho_{12} \rho_{13}-
                     2 \rho_{12} \rho_{23}-2 \rho_{13} \rho_{23}\right)}\ ,
\]
or in geometrical terms,
\[
{\tilde V_m} \ =\ \frac{3}{8}\ \frac{(m_1+m_2+m_3)}{P_m}\ +\frac{(d-2)(d-4)}{2}\
  \frac{P_m}{m_1 m_2 m_3\,S^2_{\triangle}}
\]
where the second term is absent for $d=2,4$.
The Laplace-Beltrami operator plays a role of the kinetic energy of three-dimensional quantum particle moving in curved space. It seems evident the existence of (quasi)-exactly-solvable problems with such a kinetic energy, see e.g. \cite{Crandall:1985} as for the example of exactly-solvable problem at $d=3$.


\subsection*{The symmetry operator}

For unequal masses, the symmetry operator for both  (\ref{addition3-3r-M}) and (\ref{HLB3M})
is
\[
L_{1,m} =
    \frac{\rho_{12}(m_1-m_2)+(\rho_{13}-\rho_{23})(m_1+m_2)}{m_1m_2}\pa_{\rho_{12}} + \frac{\rho_{13}(m_3-m_1)+(\rho_{23}-\rho_{12})(m_1+m_3)}{m_1m_3}\pa_{\rho_{13}}
\]
\begin{equation}
\label{Lop}
 +\ \frac{\rho_{23}(m_2-m_3)+(\rho_{12}-\rho_{13})(m_2+m_3)}{m_2m_3}\pa_{\rho_{23}}\  .
\end{equation}
which commutes,
\[
     [{\De_R}'(\rho_{ij})\ ,\ L_{1,m}]\ =\ 0\ .
\]
Invariants under this action are the functions
\[
W_1=\frac{\rho_{12}}{m_3}+\frac{\rho_{13}}{m_2}+\frac{\rho_{23}}{m_1}\ ,
\]
\[
W_2=2\sqrt{2\rho_{12}\rho_{13}+2\rho_{12}\rho_{13}+2\rho_{13}\rho_{23}
-\rho_{12}^2-\rho_{13}^2+\rho_{23}^2}\ ,
\]
\[
W_4= \rho_{23}^2-\frac{2m_1}{m_1-m_3}\rho_{13}\rho_{23}-\frac{2m_1}{m_1+m_2}\rho_{12}\rho_{13}+\frac{m_1}{m_1+m_3}\rho_{13}^2+\frac{m_1m_3}{m_2(m_1+m_3)}\rho_{13}^2\]
\[-\frac{2m_1m_2}{(m_1+m_2)(m_1+m_3)}\rho_{12}\rho_{13}
 -\frac{2m_1m_3}{(m_1+m_2)(m_1+m_3)}\rho_{12}\rho_{13}+\frac{m_1m_2}{(m_1+m_2)m_3}\rho_{12}^2 + \frac{m_1}{m_1+m_2}\rho_{12}^2\ .
\]
 These invariants are related by
\[
W_4+\frac{m_1(m_1+m_2+m_3)}{4(m_1+m_2)(m_1+m_3)}W_2^2-\frac{m_1^2m_2m_3}
{(m_1+m_2)(m_1+m_3)}W_1^2=0\ .
\]
Furthermore, all are nonnegative. In particular,
\[
W_4=(a_1\rho_{12}-a_2\rho_{13})^2+(a_3\rho_{12}-a_4\rho_{23})^2+(a_5\rho_{13}-a_6\rho_{23})^2\ ,
\]
where
\[
a_1=\frac{(m_2+m_3)m_1}{\sqrt{(m_2+m_3)\frac{m_1m_3}{m_2}}(m_1+m_2)},\
 a_2=\frac{\sqrt{(m_2+m_3)\frac{m_1m_3}{m_2}}}{m_1+m_3},\
a_3=\frac{\sqrt{(m_2+m_3)\frac{1}{m_3}}}{m_1+m_2}\ ,
\]
\[
a_4=\frac{1}{\sqrt{(m_2+m_3)\frac{1}{m3}}},\  a_5=\frac{m_1}{\sqrt{\frac{m_2}{m2+m3}}(m_1+m_3)},\ a_6=\sqrt{\frac{m_2}{m2+m3}}\ .
\]

Now we make a change of variables from $\rho_{12},\rho_{13},\rho_{23}$ to $W_1,W_3,W_4$ so that $L_1=\pa_{W_3}$ in the new coordinates. We have already defined
$W_1$ and $W_4$. While we define $W_3$ by the equations
\begin{equation}
\label{W_3}
\sqrt{W_4}\cos{\left(\frac{2\Omega\, W_3}{m_1 m_2m_3}\right)}=A(\rho_{12},\rho_{13},\rho_{23}),\
\sqrt{W_4}\cos{\left(\frac{2\Omega\, W_3}{m_1 m_2m_3}\right)}=B(\rho_{12},\rho_{13},\rho_{23})\ ,
\end{equation}
 where
\[
A= -\frac{\left(\rho_{12}m_2(m_1+m_3)-\rho_{13}m_3(m_1+m_2)\right)\Omega}{m_2m_3(m_1+m_2)(m_1+m_3)}\ ,\  \Omega=\sqrt{m_1m_2m_3(m_1+m_2+m_3)}\ ,
\]
\[
 B=-\frac{\rho_{23}(m_1+m_2)(m_1+m_3)-\rho_{12}m_1(m_1+m_3)-\rho_{13}m_1(m_1+m_2)}{(m_1+m_2)(m_1+m_3)}\ .
\]
Due to the easily verified identity
\[
W_4=A^2+B^2
\]
we see that equations (\ref{W_3}) have a locally unique solution for an angle $W_3$. In terms of these new variables we find
\begin{align*}
   {\De_R}'(\rho_{ij})\ =& \ \frac{2(m_1+m_2+m_3)W_1}{m_1m_2m_3}\pa_{W_1}^2\ +\ \frac{8m_1(m_1+m_2+m_3)W_1W_4}{(m_1+m_2)(m_1+m_3)} \pa_{W_4}^2\ +\
  \\
  & \frac{m_1^2m_2m_3W_1}{2(m_1+m_2)(m_1+m_3)W_4}\pa_{W_3}^2\ +\
   \frac{8(m_1+m_2+m_3)W_4}{m_1m_2m_3}\pa_{W_1W_4}^2\\
  &\ +\ \frac{8m_1(m_1+m_2+m_3)W_1}{(m_1+m_2)(m_1+m_3)}\pa_{W_4}
   +\ \frac{2d(m_1+m_2+m_3)}{m_1m_2m_3}\pa_{W_1}\ ,
  \\
 L_1=&\,\partial_{W_3}\ .
\end{align*}

\end{document}